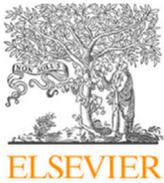
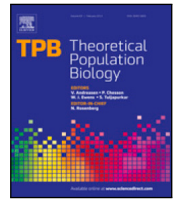

# The coalescent in finite populations with arbitrary, fixed structure

Benjamin Allen [a,*], Alex McAvoy [b,c]

[a] *Department of Mathematics, Emmanuel College, 400 The Fenway, Boston, MA, 02115, USA*
[b] *School of Data Science and Society, University of North Carolina at Chapel Hill, Chapel Hill, NC, 27599, USA*
[c] *Department of Mathematics, University of North Carolina at Chapel Hill, Chapel Hill, NC, 27599, USA*

## ARTICLE INFO

## ABSTRACT

The coalescent is a stochastic process representing ancestral lineages in a population undergoing neutral genetic drift. Originally defined for a well-mixed population, the coalescent has been adapted in various ways to accommodate spatial, age, and class structure, along with other features of real-world populations. To further extend the range of population structures to which coalescent theory applies, we formulate a coalescent process for a broad class of neutral drift models with arbitrary – but fixed – spatial, age, sex, and class structure, haploid or diploid genetics, and any fixed mating pattern. Here, the coalescent is represented as a random sequence of mappings $C = \left(C_t\right)_{t=0}^{\infty}$ from a finite set $G$ to itself. The set $G$ represents the "sites" (in individuals, in particular locations and/or classes) at which these alleles can live. The state of the coalescent, $C_t : G \to G$, maps each site $g \in G$ to the site containing $g$'s ancestor, $t$ time-steps into the past. Using this representation, we define and analyze coalescence time, coalescence branch length, mutations prior to coalescence, and stationary probabilities of identity-by-descent and identity-by-state. For low mutation, we provide a recipe for computing identity-by-descent and identity-by-state probabilities via the coalescent. Applying our results to a diploid population with arbitrary sex ratio $r$, we find that measures of genetic dissimilarity, among any set of sites, are scaled by $4r(1-r)$ relative to the even sex ratio case.

## 1. Introduction

A classical problem in population genetics is to quantify the patterns of genetic assortment that arise from a population's spatial structure (Wright, 1943; Kimura and Weiss, 1964; Maruyama, 1970; Nei, 1973; Slatkin, 1993) and mating pattern (Fisher, 1918; Wright, 1921, 1922, 1965; Allard et al., 1968; Slatkin, 1991; Caballero and Hill, 1992). Genetic assortment has important consequences for evolution, including the evolution of cooperation and other social behaviors (Hamilton, 1964; Michod, 1982; Rousset, 2004; Ohtsuki et al., 2006; Lion and van Baalen, 2008; Débarre et al., 2014; Allen et al., 2017; Hauert and Doebeli, 2021).

The coalescent (Kingman, 1982a,b; Tavaré, 2004; Rosenberg and Nordborg, 2002; Wakeley, 2009; Berestycki, 2009), a stochastic process representing ancestral lineages, is a powerful tool in this effort. Originally formulated for a well-mixed population (Kingman, 1982a,b), the coalescent has been extended to population models with spatial structure (Liggett, 1985; Cox, 1989; Notohara, 1990; Bahlo and Griffiths, 2001; Wakeley, 2001; Cox and Durrett, 2002; Nordborg and Krone, 2002; Zähle et al., 2005), uneven sex ratio (Wakeley, 2001), and polyploid genetics (Arnold et al., 2012). Many aspects of genetic assortment can be computed via the coalescent (Tavaré, 1984; Liggett, 1985; Ewens, 2004), including probabilities of identity-by-descent (Kingman, 1982b; Slatkin, 1991; Thompson, 2013).

Recently, the coalescent has emerged as an invaluable tool in evolutionary game theory (Smith and Price, 1973; Broom and Rychtár, 2013) and the evolution of social (other-affecting) behavior (Hamilton, 1964; Lehmann and Rousset, 2014; Van Cleve, 2015). Natural selection on social behavior depends critically on the patterns of genetic assortment among those performing and affected by the behavior (Hamilton, 1964; Rousset, 2004; Nowak et al., 2010; Ohtsuki, 2014). When the forces of selection are weak, these assortment patterns can be computed under neutral drift. Over time, as the mathematical theory has deepened, increasingly complex models have become tractable, from well-mixed (Nowak et al., 2004; Lessard and Ladret, 2007) to spatially homogeneous (Taylor, 1992; Rousset and Billiard, 2000; Ohtsuki et al., 2006; Taylor et al., 2007; Débarre et al., 2014) to heterogeneous (Tarnita and Taylor, 2014; Taylor and Maciejewski, 2014; Allen et al., 2017; Su et al., 2022a, 2023) populations, and from pairwise interactions to nonlinear multiplayer games (Gokhale and Traulsen, 2010; Ohtsuki, 2014; Peña et al., 2016; Archetti, 2018). In many cases, coalescent processes in finite, structured populations have enabled computation of the necessary statistics of genetic assortment to determine whether a social trait is selected (Rousset, 2003; Lessard and Ladret, 2007; Antal et al., 2009a,b; Tarnita et al., 2009; Gokhale and Traulsen, 2011; Van Cleve, 2015; Allen et al., 2017; Su et al., 2022a, 2023; Sheng et al., 2024). From these statistics, key quantities such as fixation probability under weak selection can be systematically and efficiently computed (McAvoy and Allen, 2021).

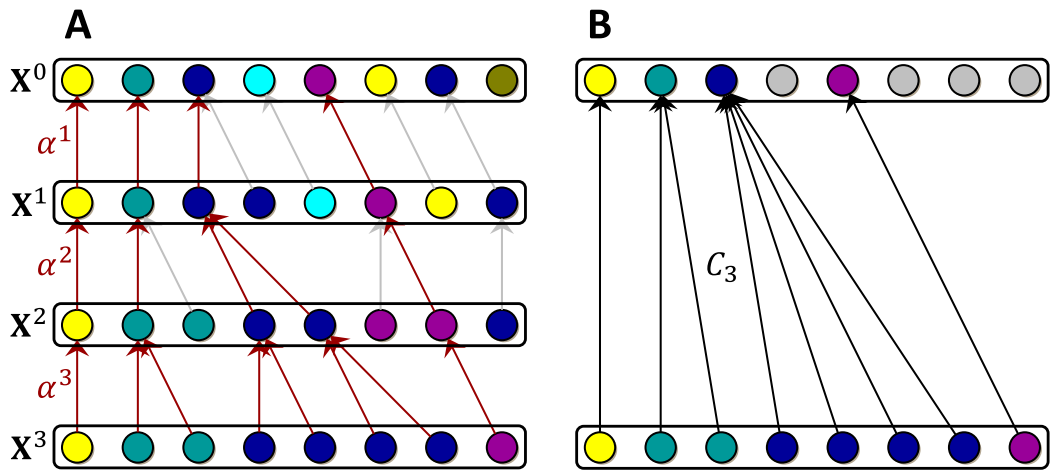

**Fig. 1. Neutral drift and coalescent mappings. A**, In the neutral drift process, $\mathcal{M} = \left(\mathbf{X}^t\right)_{t=0}^{\infty}$, each state $\mathbf{X}^t$ is identified by specifying the allele $X_g^t \in A$ (indicated by colors) occupying each genetic site $g \in G$. At each time-step, a parentage mapping $\alpha^t : G \to G$ is sampled from a fixed probability distribution, $\{p_\alpha\}_\alpha$. Offspring inherit the allele of the parent, so that $X_g^t = X_{\alpha^t(g)}^{t-1}$. **B**, The composition $C_t = \alpha^1 \circ \alpha^2 \circ \cdots \circ \alpha^t$ (shown here for $t = 3$) maps each site to the site containing the ancestor, in the initial population, of the allele occupying site $g$ at time $t$. Note that $X_g^t = X_{C_t(g)}^0$ for each site $g$. This motivates our definition of the coalescent as Markov chain $C = (C_t)_{t=0}^{\infty}$ of mappings $C_t : G \to G$ indicating where the ancestor of each present-day allele is located.

The coalescent takes different forms depending on the spatial, demographic, and/or mating structure of the model in question. Originally, Kingman (1982a,b) formulated the coalescent as a random sequence of equivalence relations representing coancestry. Modern formulations (Rosenberg and Nordborg, 2002; Wakeley, 2009) usually represent the coalescent as a process of constructing a random ancestral tree. In spatial models, the coalescent is usually represented as a collection of random walks that, upon meeting, join and step together (Holley and Liggett, 1975; Liggett, 1985; Cox, 1989; Berestycki, 2009; Barton et al., 2010).

Our aim in this work is to formulate a coalescent process for a general class of finite-population models, using a mathematical modeling framework (Allen and Tarnita, 2014; McAvoy et al., 2018; Allen and McAvoy, 2019; McAvoy and Allen, 2021) that represents structure (spatial, class, age, mating, etc.) in terms of allele transmission probabilities. Taking a gene's-eye view, this framework considers a population of alleles at a single genetic locus. Alleles live at genetic sites, within individuals, which may carry information regarding spatial location, sex, class, age, and other factors. In each transition, the alleles occupying some sites are replaced by copies of others. This framework encompasses classical models of well-mixed populations (Fisher, 1930; Wright, 1931; Moran, 1958), models with spatial (Kimura and Weiss, 1964), network (Ohtsuki et al., 2006; Taylor et al., 2007; Débarre et al., 2014; Allen et al., 2017; Su et al., 2022a), or age (Caswell, 2000) structure, as well as models with asymmetric mating patterns (Belanger et al., 2023). The key assumption is that the population size and structure remain fixed over time.

Within this class of models, we represent the coalescent as a Markov chain on mappings from the set of genetic sites to itself (Fig. 1). The state, at each time $t$, indicates the site that contained each allele's ancestor, $t$ time-steps ago. By way of these mappings, the coalescent traces the population's ancestry backwards through space and time.

The usefulness of a general framework is that any result proven from its foundational assumptions applies to all models satisfying these assumptions. The framework employed here has proven useful in deriving general mathematical results about evolutionary dynamics (Allen and Tarnita, 2014; McAvoy et al., 2018; Allen and McAvoy, 2019; McAvoy and Allen, 2021), including the evolution of social behavior (McAvoy and Wakeley, 2022; Allen et al., 2024), as well as in enabling computations for models with realistic, data-inspired features, such as heterogeneous networks (Allen et al., 2015; McAvoy et al., 2020; Su et al., 2022a,b, 2023) and hypergraphs (Sheng et al., 2024). This work aims to expand the capacity of this framework – both theoretical and computational – by furnishing it with the tools of coalescent theory.

On the theoretical side, we establish duality properties of the coalescent (Theorem 3.1) and show how the coalescent can be used to construct the stationary distribution of a neutral drift process with mutation (Section 4.5). On the computational side, we derive recurrence equations for key biological quantities: coalescence time, Eq. (3.11); coalescent branch length, Eq. (3.12); number of mutations prior to coalescence, Eq. (4.3), and identity-by-descent probability, Eq. (4.7). Perhaps our most useful result, Theorem 5.1, concerns the limit of low mutation, wherein we prove a simple relationship between expected mutation numbers and the probabilities of identity-by-descent and identity-by-state. This allows key statistics of genetic assortment, under low mutation, to be systematically computed via the coalescent. We anticipate this result being applicable to the evolution of social behavior, especially with nonlinear, multilateral interactions (Gokhale and Traulsen, 2011; Ohtsuki, 2014; Van Cleve, 2015; Allen et al., 2024; Sheng et al., 2024).

We depart from the traditional focus of coalescent theory by modeling the entire ancestry of a finite population, rather than a sample from an infinite population. This focus reflects our intended application to evolution in finite, structured populations, especially those with substantial heterogeneity. However, to connect our results to population genetics theory, we illustrate throughout how classical coalescent results can be recovered from special cases of our formalism.

We begin in Section 2 by introducing the modeling framework to which our results apply. In Section 3 we formulate the coalescent, prove duality, and analyze quantities related to reproductive value, coalescence time, branch length, and identity-by-state. Section 4 introduces mutation, demonstrates the coalescent construction of the stationary distribution, and defines identity-by-descent. The limit of low mutation is explored in Section 5. Finally, in Section 6 we apply our results to a diploid Wright–Fisher model with arbitrary sex ratio. To avoid breaking the flow of the main text, all proofs are provided in appendices. A glossary of our notation is provided in Table 1.

**Table 1**

Glossary of notation.

| Symbol | Description | Section |
|---|---|---|
| $A$ | Fixed, finite set of possible alleles | 2.1 |
| $A^G$ | Set of all $\mathbf{x} = (x_g)_{g \in G}$ with $x_g \in A$; state of $\mathcal{M}$ or $\widetilde{\mathcal{M}}$ | 2.1 |
| $\mathcal{A}$ | Markov chain on $A$ representing allele mutation | 4.2 |
| $\alpha$ | Parentage map from $G$ to itself | 2.2 |
| $C_t$ | Coalescent map at time $t$; state of $C$ | 3.1 |
| $C$ | Coalescent Markov chain, $(C_t)_{t=0}^{\infty}$ | 3.1 |
| $\widetilde{C}$ | Coalescent Markov chain with mutation, $(C_t, U_t)_{t=0}^{\infty}$ | 4.3 |
| $G$ | Fixed, finite set of genetic sites | 2.1 |
| $\iota_S(\mathbf{x})$ | Equals 1 if $x_g = x_h$ for all $g, h \in S$; 0 otherwise | 3.6 |
| $\iota_S^a(\mathbf{x})$ | Equals 1 if $x_g = a$ for all $g \in S$; 0 otherwise | 3.6 |
| $L_S$ | Branch length of coalescent tree from set $S$ in $C$ | 3.5 |
| $\ell_S$ | Expected coalescence length, $\mathbb{E}_C[L_S]$ | 3.5 |
| $M_S$ | Number of mutations in set $S$ prior to coalescence, in $\widetilde{C}$ | 4.4 |
| $m_S$ | Expected # of mutations before coalescence, $\mathbb{E}_{\widetilde{C}}[M_S]$ | 4.4 |
| $m'_S$ | $u$-derivative of $m_S$ at $u = 0$ | 5.3 |
| $\mathcal{M}$ | Markov chain on $A^G$ representing neutral drift | 2.3 |
| $\widetilde{\mathcal{M}}$ | Neutral drift Markov chain (on $A^G$) with mutation | 4.1 |
| $\mu$ | Mutant-appearance distribution (on $A^G$) | 5.2 |
| $\nu_S$ | Expected # of mutations per time-step, $\mathbb{E}_{(\alpha,S)}[|U \cap S|]$ | 4.4 |
| $\nu'_S$ | $u$-derivative of $\nu_S$ at $u = 0$ | 5.2 |
| $\pi_{\mathcal{A}}$ | Stationary distribution of alleles (i.e. of $\mathcal{A}$) | 4.2 |
| $\pi_{\widetilde{\mathcal{M}}}$ | Stationary distribution of states (i.e. of $\widetilde{\mathcal{M}}$) | 4.1 |
| $P_{\mathbf{x} \to \mathbf{y}}$ | Transition probability from $\mathbf{x}$ to $\mathbf{y}$ in $\mathcal{M}$ or $\widetilde{\mathcal{M}}$. | 2.3 |
| $p_\alpha$ | Probability of parentage map $\alpha$ | 2.2 |
| $p_{(\alpha,U)}$ | Joint probability of parentage map $\alpha$ and mutation set $U$ | 4.1 |
| $q_S$ | Stationary probability that all alleles in $S$ are IBD in $\widetilde{C}$ | 4.6.1 |
| $T_S^{\mathrm{coal}}$ | Coalescence time from set $S$ in $C$ | 3.5 |
| $\tau_S$ | Expected coalescence time, $\mathbb{E}_C[T_S^{\mathrm{coal}}]$ | 3.5 |
| $u$ | Mutation parameter | 5.1 |
| $U$ | Set of mutated sites; subset of $G$ | 4.1 |
| $v_{a \to a'}$ | Transition probability, in $\mathcal{A}$, from allele $a$ to $a'$ | 4.2 |
| $x_g$ | Allele occupying site $g \in G$ | 2.1 |
| $\mathbf{x}$ | Vector $(x_g)_{g \in G}$ of alleles occupying each site | 2.1 |
| $\mathbf{X}^t$ | State $(X_g^t)_{g \in G}$ of $\mathcal{M}$ at time $t$ | 2.1 |

## 2. Neutral drift without mutation

We begin by introducing the modeling framework within which our formulation of the coalescent is defined. This framework defines a class of models for neutral drift in structured populations. This modifies a framework introduced for other purposes by McAvoy et al. (2018), and is the neutral case of a framework developed by Allen and Tarnita (2014) and Allen and McAvoy (2019) to model natural selection. We first consider the case without mutation. Mutation will be introduced in Section 4.

### 2.1. Sites, alleles, and states

We consider neutral drift at a single genetic locus of interest. The genes at this locus belong to a finite set $A$ of possible allele types, which are considered neutral variants of each other. To represent different varieties of structure (space, group, sex, age class, etc.), we use the formalism of *genetic sites* (Allen and McAvoy, 2019). Each genetic site houses a single allele at the locus of interest. Each individual contains a number of sites equal to its ploidy: haploid individuals have one site each, diploids have two, etc. The set $G$ of all genetic sites is fixed.

For a particular model in this class, sites may be labeled with additional information. For example, in a spatial or network model, sites are associated to spatial locations or to vertices on a graph (see Section 2.4.4). For a two-sex model, sites are assigned to either male or female individuals (see Section 6). In an age-structured model, sites are labeled by age class (see Section 2.4.3). In general, however, we do not specify any explicit representation of space, sex, age, or class. Rather, this information is implicitly captured in probabilities of allele transmission, as we formalize in Section 2.2 below.

The allele occupying site $g \in G$ is recorded in a variable $x_g \in A$. The overall population state is represented by collecting all such variables into a vector $\mathbf{x} = (x_g)_{g \in G}$. The set of all possible states is denoted $A^G$.

### 2.2. Parentage maps

In a transition between states, each allele in the new state is either survived or copied from an allele in the previous state. We represent these relationships with set mappings $\alpha : G \to G$, which we call *parentage maps*. Here, $\alpha(g) \in G$ indicates the site, in the previous time-step, from which the allele currently occupying site $g \in G$ is survived or copied from. Although we refer to $\alpha$ as a "parentage map", the event $\alpha(g) = h$ can represent a variety of different biological events:

- *Haploid asexual reproduction:* $g$ is the one site in a haploid individual. This individual dies and is replaced by the clonal offspring of a parent containing site $h$.
- *Diploid sexual reproduction:* $g$ is one the two sites in a diploid individual. This individual dies and is replaced by a new offspring. Site $g$, in this new offspring, inherits the parental allele from site $h$.
- *Movement:* The individual containing site $g$ moves; correspondingly, the allele in $g$ moves to site $h$.
- *Aging:* The individual containing site $g$ progresses to a new age class; correspondingly, the allele in $g$ "moves" to site $h$.

Each of these events results in allele transmission from site $h$ to site $g$. We do not formally distinguish between these modes of allele transmission.

If parentage map $\alpha$ occurs in a given state $\mathbf{x} \in A^G$, the new state $\mathbf{x}'$ has $x'_g = x_{\alpha(g)}$ for each $g \in G$. That is, each site $g \in G$ inherits its allele from the parental site, $\alpha(g)$. We can express this compactly with the following notation: if $\gamma : G \to G$ is any set mapping, and $\mathbf{x} \in A^G$ any state, then $\mathbf{x}_\gamma \in A^G$ is the state with $(\mathbf{x}_\gamma)_g = x_{\gamma(g)}$ for all $g \in G$. With this shorthand, the new state can be denoted simply as $\mathbf{x}' = \mathbf{x}_\alpha$.

The parentage maps $\alpha$ are drawn from a fixed probability distribution $\{p_\alpha\}_\alpha$ called the *parentage distribution*. Since the alleles are considered to be neutral variants, the parentage distribution does not vary with the population state $\mathbf{x}$. The parentage distribution takes into account all aspects of spatial and mating structure, without being tied to any particular form of this structure (graphs, islands, etc.).

### 2.3. Neutral drift and fixation

To ensure unity of the population, we require that at least one genetic site be able to spread its genes throughout the population. This is formalized in the following axiom, which we require the parentage distribution $\{p_\alpha\}_\alpha$ to satisfy:

*Fixation axiom.* There exists a genetic site $g \in G$, and a finite sequence of parentage maps $\alpha_1, \ldots, \alpha_m$, with $p_{\alpha_k} > 0$ for all $1 \le k \le m$, such that $\alpha_1 \circ \alpha_2 \circ \cdots \circ \alpha_m(h) = g$ for all $h \in G$.

In words, we require there be a sequence of parentage maps, all with positive probability, leading to the outcome that all alleles are descended from a single ancestor. With this axiom, we define the neutral drift process as follows (Fig. 1A):

**Definition.** Given a parentage distribution $\{p_\alpha\}_\alpha$ satisfying the Fixation Axiom, the *neutral drift process* is a Markov chain $\mathcal{M} = (\mathbf{X}^t)_{t=0}^{\infty}$ on $A^G$. The initial state $\mathbf{X}^0 \in A^G$ is arbitrary. Each subsequent state $\mathbf{X}^{t+1}$, for $t \ge 0$, is determined from $\mathbf{X}^t$ by sampling a parentage map $\alpha$ from $\{p_\alpha\}_\alpha$ and setting $\mathbf{X}^{t+1} = \mathbf{X}_\alpha^t$ (using the shorthand notation from Section 2.2).

We denote the transition probability from $\mathbf{x}$ to $\mathbf{y}$ in $\mathcal{M}$ by $P_{\mathbf{x} \to \mathbf{y}}$.

Monoallelic states – states containing only one allele – are absorbing in $\mathcal{M}$. We denote the monoallelic state containing only allele $a \in A$ by $\mathbf{m}^a \in A^G$. Conversely, the Fixation Axiom implies that each non-monoallelic state is transient. This was proven in Theorem 2 of Allen and Tarnita (2014) for two alleles, and the proof readily extends to any finite set of alleles. This means that, with sufficient time, the population will ultimately be taken over by a single allele.

*2.4. Examples*

As we illustrate in the following examples (and in Section 6), our formalism encompasses classical population-genetic models as well as models with spatial, age, and/or class structure. The main limitation is that population size and structure must be fixed, in that the number and identity of genetic sites do not change over time.

*2.4.1. Moran process*

The Moran (1958) process models a well-mixed haploid population with overlapping generations. We confine ourselves here to the neutral case. At each time-step, a single individual is chosen uniformly at random to be replaced by the offspring of another (possibly the same) individual, also chosen uniformly at random. The parentage distribution $\{p_\alpha\}_\alpha$ is obtained by independently choosing sites $g$ and $h$ uniformly at random from $G$, and setting $\alpha(g) = h$ and $\alpha(k) = k$ for all $k \neq g$. More explicitly,

$$p_\alpha = \begin{cases} \frac{1}{N^2} & \alpha(g) \neq g \text{ for exactly one } g \in G, \\ \frac{1}{N} & \alpha = \text{Id}_G, \\ 0 & \text{otherwise}, \end{cases} \quad (2.1)$$

where $N = |G|$ is the population size. The second case arises when an allele is replaced by its own offspring.

*2.4.2. Wright-Fisher process*

The Wright–Fisher process models a well-mixed population with non-overlapping generations. We describe the haploid asexual case here; diploids with arbitrary sex ratio are analyzed in Section 6. At each time-step, each individual is replaced by the offspring of another (possibly the same) individual. Under neutrality, all such choices are uniformly random and independent of each other. The parentage distribution $\{p_\alpha\}_\alpha$ for this model is therefore the uniform distribution over all set mappings $\alpha : G \to G$.

*2.4.3. Age-structured model*

Our framework also applies to age-structured population models with a finite number of age classes of fixed size. For example, consider a haploid asexual population with two age classes, juvenile and adult. To represent these classes, we partition $G$ into disjoint sets, $G_J$ and $G_A$, for the genetic sites in juveniles and adults, respectively. Suppose that, at each time-step, some fraction of juveniles survive to adulthood, and adults produce the next generation of juveniles. The parentage map $\alpha$ then separates as the disjoint union of two maps: $\beta : G_A \to G$ indicating the previous site of each surviving individual, and $\gamma : G_J \to G_A$ indicating the parent of each new juvenile. For a well-mixed population, we may assume that $\gamma$ is sampled uniformly from all maps $G_J \to G_A$, while the distribution on $\beta$ reflects the relative likelihoods of juvenile and adult survival.

*2.4.4. death-Birth process on graphs*

The death-Birth process (Ohtsuki et al., 2006; Taylor et al., 2007; Allen et al., 2017) models a haploid, asexual, spatially structured population. The neutral case, which we confine ourselves to here, is equivalent to the well-known voter model (Clifford and Sudbury, 1973; Holley and Liggett, 1975; Cox, 1989; Sood and Redner, 2005).

For this process, $G$ is taken to be the vertex set of a weighted (undirected) graph. The edge weight between sites $g, h \in G$ denoted $w_{gh}$. The population size is $N = |G|$. At each time-step, a site $g \in G$ is chosen uniformly at random to be replaced. This individual is replaced by the offspring of a site $h \in G$, chosen with probability $p_{gh} := w_{gh} / \sum_{k \in G} w_{gk}$. Thus, the parentage distribution is obtained by (i) choosing $g$ uniformly at random from $G$; (ii) choosing $h \in G$ with probability $p_{gh}$; and (iii) setting $\alpha(g) = h$ and $\alpha(k) = k$ for all $k \neq g$.

*2.5. Symmetry*

It is helpful to introduce a formal notion of symmetry from Allen (2023). A *symmetry* of the parentage distribution $\{p_\alpha\}_\alpha$ is a permutation $\sigma : G \to G$ such that $p_\alpha = p_{\sigma^{-1} \circ \alpha \circ \sigma}$. In words, symmetries are permutations of sites that preserve the probabilities of allele transmission. For example, in the age-structured model, any permutation that preserves the partition into juvenile and adult sites is a symmetry.

The Moran and Wright–Fisher models have the property that *all* permutations are symmetries. Following Cannings (1974), Kingman (1982a), and Lessard and Ladret (2007) we refer to this property as *exchangeability*. Exchangeable models represent well-mixed populations, i.e. those without any structure relevant to the neutral drift process.

**3. The coalescent**

Having identified the relevant class of neutral drift processes, we are prepared formulate the coalescent as a Markov chain on set mappings from $G$ to itself. The relationship to other formulations of the coalescent is discussed in Section 3.2.

*3.1. Definition*

To motivate our formulation of the coalescent, consider an arbitrary initial state, $\mathbf{X}^0 = \mathbf{x}$, of $\mathcal{M}$. Suppose that the parentage maps $\alpha^1, \ldots, \alpha^T$ are sampled in successive time-steps from the parentage distribution. Then, for each $t = 1, \ldots, T$, we have $\mathbf{X}^t = \mathbf{X}^{t-1}_{\alpha^t}$ for $1 \leq t \leq T$, using the compact notation of Section 2.2 (see also Fig. 1A). Observing that this notation obeys the composition rule $(\mathbf{x}_\alpha)_\beta = \mathbf{x}_{\alpha \circ \beta}$, the final state $\mathbf{X}^T$ is related to the initial state $\mathbf{x}$ by

$$\mathbf{X}^T = \mathbf{x}_{\alpha^1 \circ \alpha^2 \circ \cdots \circ \alpha^T}. \quad (3.1)$$

The composition $C_T = \alpha^1 \circ \alpha^2 \circ \cdots \circ \alpha^T$ therefore maps each site $g \in G$ to the site containing its ancestor $T$ steps in the past (Fig. 1B). This motivates our central definition:

**Definition.** The *coalescent* $C = (C_t)_{t=0}^\infty$ is a Markov chain on set mappings from $G$ to itself. The initial state, $C_0$, is the identity mapping $\text{Id}_G$ on $G$. Each successive state $C_{t+1}$ is constructed from the previous state $C_t$ by sampling a map $\alpha$ from the parentage distribution $\{p_\alpha\}_\alpha$, and setting $C_{t+1} = \alpha \circ C_t$.

Each state $C_t$ represents a mapping from present-day alleles to their ancestors, $t$ time-steps in the past. The initial state, $C_0 = \text{Id}_G$, identifies each allele's "ancestor" in the current time-step as itself.

As a convention, we use a superscript $t$ for variables like $\mathbf{X}^t$ that move forward in chronological time as $t$ increases, and a subscript $t$ for those like $C_t$ that move backwards in chronological time with $t$.

*3.2. Partial representations of the coalescent*

A number of alternative representations of the coalescent – including classical formulations – can be obtained as functionals of the full coalescent $C$.

*3.2.1. Set of ancestors*

For each $t \geq 0$, the image $S_t := C_t(G) \subseteq G$ identifies the sites containing ancestors of the present-day population, $t$ steps into the past (Fig. 2A). The sequence $S = (S_t)_{t=0}^\infty$ is a Markov chain, with $S_{t+1}$ can be constructed from $S_t$ by sampling $\alpha$ and setting $S_{t+1} = \alpha(S_t)$. This Markov chain $S$, which we call the *coalescent on subsets*, will prove useful in obtaining recurrence relations for quantities derived from the coalescent.

The size $|S_t|$ represents the number of distinct ancestors $t$ time-steps ago. For exchangeable models (see Section 2.5), the sequence of sizes $(|S_t|)_{t=0}^\infty$ also form a Markov chain. This follows from the fact that $\mathbb{P}[|\alpha(S)| = k]$ – being equal to $\mathbb{P}[|\alpha(\sigma(S))| = k]$ for any permutation $\sigma$ of $G$ – must depend only on the size of $S$.

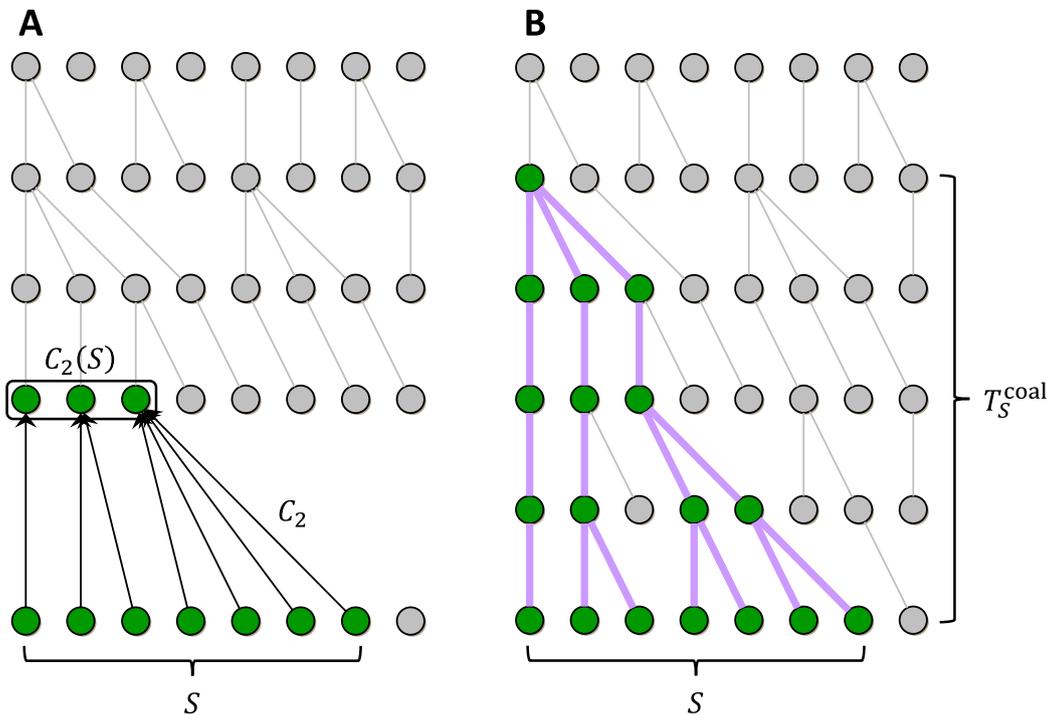

**Fig. 2. Coalescence of a subset. A**, For a subset $S \subseteq G$ representing a sample of the population, the image $C_t(S)$ (shown here for $t = 2$) represents the locations of ancestors of present-day alleles in $S$, $t$ time-steps into the past. **B**, The coalescence time from $S$, $T_S^{\text{coal}}$, is the smallest $t$ for which $|C_t(S)| = 1$. $C_{T_S^{\text{coal}}}(S)$ contains a single site $g$, representing the location of the most recent common ancestor of the alleles currently occupying $S$.

### 3.2.2. Equivalence relations

For all $t \geq 0$, we can define an equivalence relation $\sim_t$ on $G$ by $g \sim_t h$ if $C_t(g) = C_t(h)$. In words, two sites are equivalent if they have the same ancestor $t$ time-steps ago. This generalizes Kingman's (1982b, 1982a) equivalence-relation-valued coalescent to structured populations.

For exchangeable models, the sequence $(\sim_t)_{t=0}^{\infty}$ is a Markov chain. This can be seen by noting that, for any permutation $\sigma : G \to G$, and any possible realization $(c_t)_{t=0}^{T}$ of the coalescent up until time $T$,

$$\begin{aligned}
&\mathbb{P}_C\left[g \sim_{T+1} h \mid C_t = c_t, 0 \leq t \leq T\right] \\
&= \mathbb{P}_\alpha\left[\alpha \circ c_T(g) = \alpha \circ c_T(h)\right] \\
&= \mathbb{P}_\alpha\left[\sigma^{-1} \circ \alpha \circ \sigma \circ c_T(g) = \sigma^{-1} \circ \alpha \circ \sigma \circ c_T(h)\right] \\
&= \mathbb{P}_\alpha\left[\alpha \circ \sigma \circ c_T(g) = \alpha \circ \sigma \circ c_T(h)\right].
\end{aligned} \quad (3.2)$$

Because this equation holds for any permutation $\sigma$, the value of the two sides depends only on whether $c_T(g) = c_T(h)$ – i.e. on whether $g \sim_T h$ – and not otherwise on any values of $(c_t)_{t=0}^{T}$. Therefore, $\sim_{T+1}$ is conditionally independent of $(C_t)_{t=0}^{T}$ given $\sim_T$, verifying the Markov property for $(\sim_t)_{t=0}^{\infty}$.

### 3.2.3. Coalescent restricted to a subset

For any nonempty subset $S \subseteq G$, let $C_{t|S} : S \to G$ denote the restriction of $C_t$ to $S$. Then $C_{|S} = (C_{t|S})_{t=0}^{\infty}$ is also a Markov chain, representing the ancestry of set $S$. This corresponds to the usual focus of coalescent theory on the ancestry of a sample from a population (Kingman, 1982b; Rosenberg and Nordborg, 2002; Wakeley, 2009).

### 3.2.4. Coalescing random walks

In spatial environments, the coalescent is often formulated as a collection of random walks that step together after meeting (Holley and Liggett, 1975; Liggett, 1985; Cox, 1989; Berestycki, 2009). This perspective is recovered in our formalism by defining $Y_t^g = C_t(g) \in G$ for each $g \in G$ and $t \geq 0$. Each sequence $(Y_t^g)_{t \geq 0}$ can be considered as a random walk on $G$. We observe that any two such walks, $(Y_t^g)_{t \geq 0}$ and

$(Y_t^h)_{t \geq 0}$, step together after meeting. The collection of all such walks $\left((Y_t^g)_{g \in G}\right)_{t \geq 0}$ is equivalent to $C$ itself.

Coalescing random walk theory typically assumes that the walkers step independently before meeting. However, this is not necessarily true in our framework: $Y_{t+1}^g$ and $Y_{t+1}^h$ need not be independent conditioned on distinct values of $Y_t^g$ and $Y_t^h$. Such an independence property only holds if $\alpha(g)$ is independent of $\alpha(h)$, for all $g, h \in G$ with $h \neq g$, under $\{p_\alpha\}_\alpha$. The Wright–Fisher process, for example, satisfies this independence property. A similar independence property arises for models in which exactly one site is replaced per time-step – such as the Moran and death-Birth processes – but only upon replacing the discrete-time walks $(Y_t^g)_{t \geq 0}$ with their continuous-time analogues.

### 3.3. Coalescent duality

A key principle of coalescent theory is a duality between the forwards-time neutral drift process and the backwards-time coalescent process (Holley and Liggett, 1975; Liggett, 1985; Berestycki, 2009). This duality arises from the fact that, absent mutation, each allele in the current population is a copy of its ancestor in the initial population.

To formalize the duality principle for our coalescent, we introduce a coupling of $C$ with the neutral drift Markov chain, $\mathcal{M}$. The coupled process, denoted $(\mathcal{M}, C)_{t=0}^{T}$, consists of a finite random sequence of pairs $(\mathbf{X}^t, C_t)_{t=0}^{T}$, with $\mathbf{X}^t \in A^G$ and $C_t : G \to G$ for $0 \leq t \leq T$. Given an arbitrary initial state $\mathbf{x}$, the coupled process $(\mathcal{M}, C)_{t=0}^{T}$ is constructed as follows:

1. Sample a sequence of parentage maps $\alpha^1, \ldots, \alpha^T$ independently from the parentage distribution $\{p_\alpha\}_\alpha$;
2. Set $\mathbf{X}^0 = \mathbf{x}$ and $\mathbf{X}^t = \mathbf{X}_{\alpha^t}^{t-1}$ for each $t = 1, \ldots, T$;
3. Set $C_0 = \text{Id}_G$ and $C_{t+1} = \alpha^{T-t} \circ C_t$ for each $t = 1, \ldots, T$.

Above, the $\alpha^t$'s are given superscript indices to indicate that they proceed forward in time; e.g. $\alpha^1$ occurs first chronologically. Correspondingly, the sequence $(\mathbf{X}^t)_{t=0}^{T}$ moves forwards in chronological time from initial state $\mathbf{X}^0 = \mathbf{x}$ to final state $\mathbf{X}^T$, while $(C_t)_{t=0}^{T}$ proceeds backwards in chronological time with $t$. It is because of these opposite-time orientations that the coupled process $(\mathcal{M}, C)_{t=0}^T$ is defined only for finite time; there does not appear to be an obvious infinite-time extension.

From the definition of $(\mathcal{M}, C)_{t=0}^T$, we obtain the identity

$$\mathbf{X}^T = \mathbf{X}_{C_t}^{T-t}, \tag{3.3}$$

for each $t = 0, \ldots, T$. In particular, setting $t = T$,

$$\mathbf{X}^T = \mathbf{x}_{C_T}. \tag{3.4}$$

Eq. (3.4) encapsulates the duality principle: Since the occupant of each site $g \in G$ at time $t = T$ is descended (without mutation) from the occupant of site $C_T(g)$ at time $t = 0$, they must have the same allele, $X_g^T = x_{C_T(g)}$. Therefore, for any arbitrary states $\mathbf{x}, \mathbf{y} \in A^G$, we have

$$\mathbb{P}_{(\mathcal{M}, C)_{t=0}^T} \left[ \mathbf{X}^T = \mathbf{y} \mid \mathbf{X}^0 = \mathbf{x} \right] = \mathbb{P}_{(\mathcal{M}, C)_{t=0}^T} \left[ \mathbf{x}_{C_T} = \mathbf{y} \right]. \tag{3.5}$$

Now "forgetting" the coupling on each side, we arrive at the formal statement of coalescent duality:

**Theorem 3.1** (*Coalescent Duality*). *If $\mathbf{x}$ is an arbitrary initial state, then, for any time $T \geq 0$ and state $\mathbf{y} \in A^G$,*

$$\mathbb{P}_{\mathcal{M}} \left[ \mathbf{X}^T = \mathbf{y} \mid \mathbf{X}^0 = \mathbf{x} \right] = \mathbb{P}_C \left[ \mathbf{x}_{C_T} = \mathbf{y} \right]. \tag{3.6}$$

As a consequence, for any state function $f : A^G \to \mathbb{R}$, initial state $\mathbf{x}$, and time $t \geq 0$, we have

$$\mathbb{E}_{\mathcal{M}} \left[ f(\mathbf{X}^t) \mid \mathbf{X}^0 = \mathbf{x} \right] = \mathbb{E}_C \left[ f\left(\mathbf{x}_{C_t}\right) \right]. \tag{3.7}$$

### 3.4. Stationary distribution of the coalescent

Ultimately, tracing back far enough in the past, all alleles in the present-day population must be descended from a common ancestor. This is represented by the state of the coalescent being equal to a constant map on $G$. Specifically, letting $\kappa_g$ denote the constant map to $g \in G$ (that is, $\kappa_g(h) = g$ for all $h \in G$), the event $C_t = \kappa_g$ indicates there was common ancestor, at site $g$, $t$ time-steps ago. Since all nonconstant maps are transient in $C$, this is guaranteed to occur eventually.

Moreover, we prove in Theorem B.1 that, as $t \to \infty$, $C$ converges to a unique stationary distribution on these constant maps. In this stationary distribution, the probabilities, $\rho_g$, of each constant map, $\kappa_g$, are uniquely determined by the system of equations

$$\rho_g = \sum_\alpha p_\alpha \sum_{h \in \alpha^{-1}(g)} \rho_h; \tag{3.8a}$$

$$\sum_{g \in G} \rho_g = 1. \tag{3.8b}$$

The stationary probability $\rho_g$ is also equal to the probability that a neutral mutation, arising at site $g$, ultimately becomes fixed in the population. Indeed, it was shown in other works (Maciejewski, 2014; Allen et al., 2015; Allen and McAvoy, 2019) that these neutral fixation probabilities are uniquely determined by Eq. (3.8). This can also be shown using coalescent duality. In this way, $\rho_g$ can be understood as the *reproductive value* of site $g$, connecting to the classical concept of reproductive value (Fisher, 1930; Taylor, 1990; Barton and Etheridge, 2011) as the expected contribution to the future gene pool under neutral drift. For example, in the death-Birth process, solving Eq. (3.8) yields that each site $g$ has reproductive value proportional to the sum of adjacent edge weights: $\rho_g = \sum_{h \in G} w_{gh} / \sum_{h,k \in G} w_{kh}$ (Maciejewski, 2014; Allen and McAvoy, 2019).

### 3.5. Coalescence

The event that $C_t$ is constant for the first time is called *coalescence*; this represents the meeting of lineages in the most recent common ancestor. For a subsets $S$ of sites, *coalescence of $S$* is the event that $C_t(g) = C_t(h)$, for all $g, h \in S$, for the first time (Fig. 2B).

Two fundamental quantities of interest are the time to coalescence (Fig. 2B), and the total branch length of the coalescent tree, for a given set $S$:

**Definition.** For any nonempty $S \subseteq G$, the *coalescence time* from $S$ is

$$T_S^{\text{coal}} := \min \left\{ t \geq 0 \,:\, |C_t(S)| = 1 \right\}, \tag{3.9}$$

and the *coalescence length* from $S$ is

$$L_S := \sum_{t=0}^{T_S^{\text{coal}} - 1} |C_t(S)|. \tag{3.10}$$

Since all nonconstant mappings are transient in $C$, $T_S^{\text{coal}}$ has an geometric-tailed distribution for each $S$ (Kemeny and Snell, 1976, Corollary 3.1.2). It follows that all moments of $T_S^{\text{coal}}$ are finite, and the same is true of $L_S$, since $L_S \leq |S| T_S^{\text{coal}}$. We denote the expectations of $T_S^{\text{coal}}$ and $L_S$ by $\tau_S$ and $\ell_S$, respectively. We prove in Theorem C.1 that $\tau_S$ and $\ell_S$ are uniquely determined by the recurrence relations

$$\tau_S = \begin{cases} 1 + \sum_\alpha p_\alpha \tau_{\alpha(S)} & |S| \geq 2, \\ 0 & |S| = 1, \end{cases} \tag{3.11}$$

and

$$\ell_S = \begin{cases} |S| + \sum_\alpha p_\alpha \ell_{\alpha(S)} & |S| \geq 2, \\ 0 & |S| = 1. \end{cases} \tag{3.12}$$

Theorem C.1 also provides recurrence equations for higher-order moments of $T_S^{\text{coal}}$ and $L_S$, allowing them to be iteratively computed from moments of lower order.

**Example 3.1.** For the Moran process (Section 2.4.1), exchangeability implies that the distributions of $T_S^{\text{coal}}$ and $L_S$ depend only on the size of $S$. For a set $S$ of size $k$, the image $\alpha(S)$ has size $k - 1$ with probability $k(k-1)/N^2$, and otherwise remains size $k$. Consequently, the recurrence relations (3.11) and (3.12) have the form

$$\psi_k = \begin{cases} f(k) + \left(1 - \dfrac{k(k-1)}{N^2}\right) \psi_k + \dfrac{k(k-1)}{N^2} \psi_{k-1} & k \geq 2, \\ 0 & k = 1. \end{cases} \tag{3.13}$$

For the expected coalescence times $\tau_k$, $f(1) = 0$ and $f(k) = 1$ for $k \geq 2$. For expected coalescence lengths $\ell_S$, $f(1) = 0$ and $f(k) = k$ for $k \geq 2$. The general solution to Eq. (3.13) is

$$\psi_k = N^2 \sum_{j=1}^{k-1} \frac{f(j+1)}{j(j+1)}. \tag{3.14}$$

Substituting the appropriate functions $f(k)$ yields

$$\tau_k = N^2 \sum_{j=1}^{k-1} \frac{1}{j(j+1)} = N^2 \left(1 - \frac{1}{k}\right); \tag{3.15a}$$

$$\ell_k = N^2 \sum_{j=1}^{k-1} \frac{1}{j}, \tag{3.15b}$$

in accordance with standard results (Watterson, 1975; Hudson, 1990; Donnelly and Tavare, 1995).

**Example 3.2.** For the death-Birth process on graphs (Section 2.4.4), Eq. (3.11) for expected coalescence times becomes

$$\tau_S = \begin{cases} \frac{N}{|S|} + \frac{1}{|S|} \sum_{g \in S} \sum_{h \in G} p_{gh}\, \tau_{(S \setminus \{g\}) \cup \{h\}} & |S| \geq 2, \\ 0 & |S| = 1, \end{cases} \quad (3.16)$$

and Eq. (3.12) for coalescence lengths becomes

$$\ell_S = \begin{cases} N + \frac{1}{|S|} \sum_{g \in S} \sum_{h \in G} p_{gh}\, \ell_{(S \setminus \{g\}) \cup \{h\}} & |S| \geq 2, \\ 0 & |S| = 1. \end{cases} \quad (3.17)$$

*3.6. Identity by state*

One of the primary applications of coalescent theory is to compute the probability that the same allele co-occurs in a given set of individuals or sites. In a given state **x** of $\mathcal{M}$, we say a nonempty subset $S \subseteq G$ is *identical-by-state (IBS)* if all sites in $S$ contain the same allele. We quantify identity-by-state using an indicator function $\iota_S : A^G \to \{0,1\}$, equal to 1 if $S$ is IBS and 0 otherwise:

$$\iota_S(\mathbf{x}) := \begin{cases} 1 & \text{there exists } a \in A \text{ such that } x_g = a \text{ for all } g \in S, \\ 0 & \text{otherwise.} \end{cases} \quad (3.18)$$

These IBS indicator functions obey $\iota_S(\mathbf{x}_\beta) = \iota_{\beta(S)}(\mathbf{x})$ for any set mapping $\beta : G \to G$. Since all non-monoallelic states are transient, we have $\lim_{t\to\infty} \mathbb{E}_{\mathcal{M}}[\iota_S(\mathbf{X}^t)] = 1$, regardless of the initial state of $\mathcal{M}$.

It is also useful to quantify whether a set $S$ contains only a particular allele $a \in A$. We do this with an indicator function $\iota_S^a : A^G \to \{0,1\}$, defined by

$$\iota_S^a(\mathbf{x}) := \begin{cases} 1 & x_g = a \text{ for all } g \in S, \\ 0 & \text{otherwise.} \end{cases} \quad (3.19)$$

A useful measure of genetic dissimilarity within a nonempty subset $S \subseteq G$ is the duration of time, in $\mathcal{M}$, for which $S$ is *not* IBS:

$$T_{S|\mathbf{x}}^{\text{IBS}} := \sum_{t=0}^{\infty} \left(1 - \iota_S(\mathbf{X}^t)\right). \quad (3.20)$$

The distribution of $T_S^{\text{IBS}}$ depends on the initial state $\mathbf{X}^0$ of $\mathcal{M}$. By coalescent duality (Theorem 3.1), we have

$$\mathbb{E}_{\mathcal{M}}\left[T_S^{\text{IBS}} \mid \mathbf{X}^0 = \mathbf{x}\right] = \sum_{t=0}^{\infty} \mathbb{E}_C\left[1 - \iota_S\left(\mathbf{x}_{C_t}\right)\right] = \sum_{t=0}^{\infty} \mathbb{E}_C\left[1 - \iota_{C_t(S)}(\mathbf{x})\right]. \quad (3.21)$$

In particular, suppose **x** is a state for which all sites have different alleles, $x_g \neq x_h$ for all $h \neq g$. Then $\iota_{C_t(S)} = 1$ if and only if $C_t$ is a constant map, and it follows from Eq. (3.21) that $\mathbb{E}\left[T_S^{\text{IBS}} \mid \mathbf{X}^0 = \mathbf{x}\right] = \tau_S$. In the case $S = G$, $T_{G|\mathbf{x}}^{\text{IBS}}$ and $T_G^{\text{coal}}$ are equal not only in expectation but in their full distributions. This can be seen using a variation of the coupled process $(\mathcal{M}, C)_{t=0}^T$ that halts at $T = T_G^{\text{IBS}}$ instead of at a predetermined $T$; such a process will have $T_G^{\text{coal}} = T_G^{\text{IBS}}$ as a consequence of Eq. (3.3). In words, the coalescence time for the entire population equals the time to fixation from an initial state with all different alleles.

## 4. The coalescent with mutation

We now augment the neutral drift process, $\mathcal{M}$, and coalescent, $C$, to incorporate mutation.

*4.1. Mutation sets*

We suppose that, in each time-step of the neutral drift process, some (possibly empty) subset $U \subseteq G$ of sites acquire mutations. A transition is represented by the pair $(\alpha, U)$. The probability that transition $(\alpha, U)$ occurs in a particular time-step is given by a fixed probability distribution $\{p_{(\alpha,U)}\}_{(\alpha,U)}$.

The only requirement we place on the joint distribution of $\alpha$ and $U$ is that it be possible for a new mutation to sweep to fixation. To formalize this, we amend the Fixation Axiom to ensure that there is a site from which mutation and subsequent fixation is possible:

*Fixation axiom with mutation.* There exists a genetic site $g \in G$ with $\mathbb{P}_{(\alpha,U)}[g \in U] > 0$, and a finite sequence of parentage maps $\alpha_1, \ldots, \alpha_m$ with $p_{(\alpha_k,\emptyset)} > 0$ for all $1 \leq k \leq m$, such that $\alpha_1 \circ \alpha_2 \circ \cdots \circ \alpha_m(h) = g$ for all $h \in G$.

We take this amended axiom to be in force for the remainder of this section.

**Example 4.1.** Mutation is introduced in the death-Birth process (Section 2.4.4) by supposing that each new offspring acquires a mutation with a given probability $0 < u < 1$ (Allen et al., 2012). Formally, $\alpha$ and $U$ are jointly constructed by (i) choosing $g \in G$ uniformly at random, (ii) choosing $h \in G$ with probability $p_{gh}$, (iii) setting $\alpha(g) = h$ and $\alpha(k) = k$ for all $k \neq g$, and (iv) setting $U = \{g\}$ with probability $u$ and $U = \emptyset$ otherwise. The Moran process arises as a special case with $p_{gh} = \frac{1}{N}$ for all $g, h \in G$.

**Example 4.2.** In the Wright–Fisher process (Section 2.4.2), mutations occur with a given probability $0 < u < 1$ in each offspring, independently of all others. Formally, each set $U \subseteq G$ is chosen with probability $u^{|U|}(1-u)^{|G|-|U|}$, independently of $\alpha$.

*4.2. Mutation of alleles*

We represent mutation between alleles as a (finite) Markov chain, $\mathcal{A}$, on $A$. The transition probability from $a \in A$ to $a' \in A$ (representing the probability that mutation of $a$ results in $a'$) is denoted $v_{a \to a'}$.

In order to guarantee a unique stationary distribution for neutral drift with mutation, we assume that $\mathcal{A}$ has a single closed communicating class. This means that there exists a subset of alleles that are accessible (via a finite sequence of transitions with positive probability) from an arbitrary initial allele. With this assumption, the Perron–Frobenius Theorem guarantees that $\mathcal{A}$ has a unique stationary distribution, $\pi_{\mathcal{A}}$, which is zero outside of this closed communicating class. (We do not require that $\mathcal{A}$ be aperiodic. Indeed, periodic cases may be of interest, such as mutation that interchanges two alleles.)

For some results, we will also assume there is no self-mutation of alleles; i.e. $v_{a \to a} = 0$ for each $a \in A$. This assumption will be used to establish links between identity-by-descent (introduced in Section 4.6) and identity-by-state.

Given a probability distribution, $\{p_{(\alpha,U)}\}_{(\alpha,U)}$ and a mutation Markov chain, $\mathcal{A}$, satisfying the above assumptions, a Markov chain $\widetilde{\mathcal{M}} = (\mathbf{X}^t)_{t=0}^{\infty}$, representing neutral drift with mutation, is defined as follows. Given an initial state, $\mathbf{X}^0 = \mathbf{x}$, each successive state, $\mathbf{X}^{t+1}$, is constructed from $\mathbf{X}^t$ by sampling $(\alpha, U)$ from $\{p_{(\alpha,U)}\}_{(\alpha,U)}$ and

- for each non-mutated site, $g \in G \setminus U$, setting $X_g^{t+1} = X_{\alpha(g)}^t$;
- for each mutated site $g \in U$, independently sampling $X_g^{t+1}$ from the probability distribution $\left\{v_{X_{\alpha(g)}^t \to a}\right\}_{a \in A}$.

We prove in Lemma D.1 that $\widetilde{\mathcal{M}}$ converges to a unique stationary distribution. A result of McAvoy et al. (2018) implies that, in this stationary distribution, the marginal distribution for the allele at any given site is $\pi_{\mathcal{A}}$:

$$\mathbb{P}_{\pi_{\widetilde{\mathcal{M}}}}\left[X_g = a\right] = \pi_{\mathcal{A}}(a) \quad \text{for each } g \in G \text{ and } a \in A. \quad (4.1)$$

*4.3. Coalescent with mutation*

Given a probability distribution $\{p_{(\alpha,U)}\}_{(\alpha,U)}$ satisfying the Fixation Axiom with Mutation, we define the coalescent with mutation as follows:

**Definition.** The *coalescent with mutation* is a Markov chain, $\widetilde{C} = (C_t, U_t)_{t=0}^{\infty}$, with $C_t : G \to G$ and $U_t \subseteq G$, constructed as follows: First, set $C_0 = \mathrm{Id}_G$. Then, for each $t \geq 0$, sample $(\alpha, U)$ from $\{p_{(\alpha,U)}\}_{(\alpha,U)}$, and set $U_t = U$ and $C_{t+1} = \alpha \circ C_t$.

In this definition, $U_t$ represents the set of sites that acquired mutations $t$ time-steps in the past. This motivates the unusual time offset, in which $U_t$ and $C_{t+1}$ are determined simultaneously.

*4.4. Mutations prior to coalescence*

We denote the expected number of mutations that set $S$ accrues in a single time-step by $\nu_S := \mathbb{E}_{(\alpha,U)}[|U \cap S|]$. For singleton sets, we abbreviate $\nu_{\{g\}}$ to $\nu_g$. By additivity of expectations, $\nu_S = \mathbb{E}_{(\alpha,U)}\left[\sum_{g \in S}|U \cap \{g\}|\right] = \sum_{g \in S} \nu_g$.

We are particularly interested in the number, $M_S$ of mutations in the ancestry of a nonempty set $S \subseteq G$ prior to coalescence:

$$M_S := \sum_{t=0}^{T_S^{\mathrm{coal}}-1} |U_t \cap C_t(S)|. \qquad (4.2)$$

If $S$ is a singleton set, the sum is empty, and thus $M_S = 0$. We denote the expected number of mutations before coalescence by $m_S := \mathbb{E}_{\widetilde{C}}[M_S]$. These are uniquely determined to the system of recurrence equations (Theorem C.1)

$$m_S = \begin{cases} \nu_S + \sum_\alpha p_\alpha m_{\alpha(S)} & |S| \geq 2, \\ 0 & |S| = 1. \end{cases} \qquad (4.3)$$

Iterating this equation leads to $m_S = \mathbb{E}_C\left[\sum_{t=0}^{T_S^{\mathrm{coal}}-1} \nu_{C_t(S)}\right]$. In particular, if mutation is uniform over sites, in the sense that $\nu_g = \nu$ for some $\nu > 0$ and each $g \in G$, then $m_S = \nu \ell_S$. In general, $m_S$ can be understood as a rescaling of the expected coalescence length $\ell_S$ by the site-specific mutation rates $\nu_g$.

*4.5. Constructing the stationary distribution on $\widetilde{\mathcal{M}}$*

The coalescent with mutation leads to a recipe for constructing the stationary distribution $\pi_{\widetilde{\mathcal{M}}}$. The idea is to sample $\widetilde{C}$ until a common ancestor is reached, randomly assign an allele to that ancestor from the stationary distribution on $\mathcal{A}$, and then carry this allele forward to the present-time population, resolving any mutations according to $\mathcal{A}$. Similar constructions are often used for the classical coalescent process in a well-mixed population (Griffiths and Tavaré, 1994a,b). We formalize this construction as follows:

1. Sample the coalescent with mutation, $\widetilde{C}$, until time $\widehat{T} = T_G^{\mathrm{coal}}$. Let $h$ be the (single) element of $C_{\widehat{T}}(G)$, so that $C_{\widehat{T}}(g) = h$ for all $g \in G$.
2. Sample an allele $Y_h^0 \in A$ from $\pi_{\mathcal{A}}$.
3. Iteratively determine a random sequence $(\mathbf{Y}^t)_{t=1}^{\widehat{T}}$ of vectors of alleles, with each $\mathbf{Y}^t = (Y_g^t)_{g \in C_{\widehat{T}-t}(G)} \in A^{C_{\widehat{T}-t}(G)}$, such that, for each $t = 1, \ldots, \widehat{T}$ and $g \in G$,

   - If $C_{\widehat{T}-t}(g) \notin U_{\widehat{T}-t}$, then $Y_{C_{\widehat{T}-t}(g)}^t = Y_{C_{\widehat{T}-t+1}(g)}^{t-1}$,
   - If $C_{\widehat{T}-t}(g) \in U_{\widehat{T}-t}$, then $Y_{C_{\widehat{T}-t}(g)}^t = a$ with probability $\nu_{Y_{C_{\widehat{T}-t+1}(g)}^{t-1} \to a}$.

4. Set $\mathbf{X} = \mathbf{Y}^{\widehat{T}}$, noting that $\mathbf{Y}^{\widehat{T}} \in A^G$ since $C_0(G) = G$.

We prove in Theorem D.2 that this indeed results in the stationary distribution on $\widetilde{\mathcal{M}}$. It is also useful to consider the pairing,

$$\left(\widetilde{C}; \mathbf{X}\right) = \left((C_0, U_0), (C_1, U_1), \ldots; \mathbf{X}\right), \qquad (4.4)$$

comprising the coalescent with mutation, $\widetilde{C}$, together with the state, $\mathbf{X}$, constructed as above. The following properties arise directly from the construction of this pairing:

**Corollary 4.1.** *In the pairing $\left(\widetilde{C}; \mathbf{X}\right)$, for each nonempty $S \subseteq G$,*

- $\mathbb{E}\left[\iota_S \mid M_S = 0\right] = 1$;
- $\mathbb{E}\left[\iota_S^a \mid M_S = 0\right] = \pi_{\mathcal{A}}(a)$ for each $a \in A$;
- *If there is no self-mutation, then* $\mathbb{E}\left[\iota_S \mid M_S = 1\right] = 0$;
- *The probability that $(X_g)_{g \in S}$ contains alleles $a, a' \in A$, conditional on $M_S = 1$, is $\pi_{\mathcal{A}}(a) v_{a \to a'} + \pi_{\mathcal{A}}(a') v_{a' \to a}$.*

The above construction can also be applied to a subset $S$ of sites, yielding the marginal stationary distribution of alleles in $S$. That is, for fixed nonempty $S \subseteq G$, one can sample the restricted coalescent $\widetilde{C}_{|S} = (C_{t|S}, U_t \cap C_t(S))_{t=0}^{\infty}$ until time $T_S^{\mathrm{coal}}$, sample an initial allele $Y_h^0 \in A$ from $\pi_{\mathcal{A}}$ (where $h$ is the single element of $C_{T_S^{\mathrm{coal}}}(S)$), and carry mutations forward in the manner of step 3 above. An argument similar to the proof of Theorem D.2 shows that distribution on the resulting state, $\mathbf{Y}^{T_S^{\mathrm{coal}}}$, is equal to the marginal distribution on $\mathbf{X}_{|S} := (X_g)_{g \in S}$, with $\mathbf{X}$ sampled from $\pi_{\widetilde{\mathcal{M}}}$.

*4.6. Identity by descent*

Identity-by-descent (Cotterman, 1940; Malécot, 1948; Jacquard, 1974; Thompson, 2013) is a central concept and useful computational tool in population genetics. A set of alleles are *identical by descent* if no mutation separates any of them from their common ancestor.

*4.6.1. Definition*

We represent identity-by-descent as a random equivalence relation on $G$, constructed from the coalescent with mutation, $\widetilde{C}$:

**Definition.** In $\widetilde{C}$, sites $g, h \in G$ are *identical-by-descent (IBD)*, denoted $g \sim h$, if $U_t \cap C_t(\{g, h\}) = \varnothing$ for all $t \leq T_{\{g,h\}}^{\mathrm{coal}}$. Equivalently, $g \sim h$ if and only if $C_t(g) = C_t(h)$ whenever $C_t(g) \in U_t$ or $C_t(h) \in U_t$.

In words, $g$ and $h$ are IBD if they are descended from a common allele with no intervening mutation. That IBD is an equivalence relation follows directly from the second form of the definition. Consequently, there exists a (random) partition of $G$ into equivalence classes, which we call *IBD classes*. It is straightforward to show that a nonempty set $S$ is contained within a single IBD class if and only if $M_S = 0$.

*4.6.2. IBD probability*

We are particularly interested in the probability that a given set of sites are identical-by-descent:

**Definition.** The *IBD probability* of a set $S \subseteq G$ is

$$q_S := \mathbb{P}_{\widetilde{C}}\left[g \sim h \text{ for all } g, h \in S\right] = \mathbb{P}_{\widetilde{C}}\left[M_S = 0\right]. \qquad (4.5)$$

An alternative characterization of $q_S$ is obtained by considering a Markov chain $S^*$, which is a variation on the "set of ancestors" chain $S$ introduced in Section 3.2.1. The state space of $S^*$ consists of nonempty subsets of $G$ together with a symbol, $\mathcal{J}$, representing mutation prior to coalescence. Each state $S_{t+1}^*$ is constructed from the previous state, $S_t^*$, by sampling $(\alpha, U)$ from $\{p_{(\alpha,U)}\}_{(\alpha,U)}$ and setting

$$S_{t+1}^* = \begin{cases} \mathcal{J} & S_t^* = \mathcal{J}, \text{ or } |S_t^*| \geq 2 \text{ and } U \cap S_t^* \neq \varnothing, \\ \alpha(S_t^*) & \text{otherwise.} \end{cases} \qquad (4.6)$$

The Markov chain $S^*$ has two absorbing classes: one containing only state $\mathcal{J}$; the other containing singleton sets $\{g\}$ for sites $g$ satisfying the conditions of the Fixation axiom. The IBD probability $q_S$ equals the probability of absorption in the latter class, from initial state $S_0^* = S$.

From the standard recurrence relations for absorption probabilities (e.g. Theorem 3.3.7 of Kemeny and Snell, 1976), we obtain that the $q_S$ are uniquely determined by

$$q_S = \begin{cases} \sum_{\substack{(\alpha,U) \\ U \cap S = \emptyset}} p_{(\alpha,U)} q_{\alpha(S)} & |S| \geq 2, \\ 1 & |S| = 1. \end{cases} \quad (4.7)$$

Eq. (4.7) generalizes recurrence relations for IBD probabilities that have been derived for a variety of models (Cotterman, 1940; Malécot, 1948; Harris, 1964; Cockerham, 1971; Jacquard, 1974).

**Example 4.3.** For the Moran process, solving Eq. (4.7) gives

$$q_S = \prod_{k=1}^{|S|-1} \frac{(1-u)\left(\frac{k}{N}\right)}{1-(1-u)\left(1-\frac{k}{N}\right)} = \prod_{k=1}^{|S|-1} \frac{k}{\theta+k}, \quad \text{where } \theta = \frac{Nu}{1-u}.$$

This recovers a special case of the Ewens Sampling Formula (Ewens, 1972) – in the exact form derived by Kelly (1977) – for the probability that all alleles in a sample are IBD.

**Example 4.4.** For the death-Birth process, Eq. (4.7) for IBD probabilities becomes

$$q_S = \begin{cases} \dfrac{1-u}{|S|} \sum_{g \in S} \sum_{h \in G} p_{gh} q_{(S \setminus \{g\}) \cup \{h\}} & |S| \geq 2, \\ 1 & |S| = 1. \end{cases} \quad (4.8)$$

A special case arises for models, such as the Wright–Fisher model (Section 2.4.2), in which mutation occurs with a fixed probability $u$ at each site, independently of other sites and of $\alpha$. In this case, Eq. (4.7) can be solved in terms of the coalescent process $\widetilde{C}$:

$$q_S = \mathbb{E}_{\widetilde{C}}\left[\prod_{t=0}^{T^{\text{coal}}_S - 1} (1-u)^{|C_t(S)|}\right] = \mathbb{E}_{\widetilde{C}}\left[(1-u)^{L_S}\right], \quad (4.9)$$

where $L_S$ is the coalescence length defined in Eq. (3.10). This generalizes Slatkin's (1991) observation that, for pairs $g, h \in G$, $q_{\{g,h\}} = \mathbb{E}_{\widetilde{C}}\left[(1-u)^{2T_{\{g,h\}}}\right]$ (note that $L_{\{g,h\}} = 2T_{\{g,h\}}$). Eq. (4.9) has a simple intuition: under the above assumptions, each unit of branch length represents an independent chance of mutation, so the probability of none prior to coalescence is $(1-u)^{L_S}$.

## 5. The low-mutation limit

The low-mutation limit is relevant to many biological settings in which mutation in loci of interest is rare. It is also of theoretical interest, in that it simplifies a number of relationships, and links the mutation-free results of Sections 2 and 3 with those of Section 4. Our main result of this section, Theorem 5.1, shows that quantities regarding identity-by-state, identity-by-descent, and expected mutation numbers become asymptotically equivalent under low mutation.

### 5.1. Assumptions

To study the low-mutation limit, we introduce a mutation parameter, $u$, taking values in the interval $[0, \varepsilon)$ for some $\varepsilon > 0$. This mutation parameter has no fixed intrinsic meaning. It may represent, for example, the probability of mutation per time-step or per birth. We consider a family of transition rules $\{p_{(\alpha,U)}\}_{(\alpha,U)}^u$, indexed by $u \in [0, \varepsilon)$, subject to the following assumptions:

*Assumptions for low mutation.* The family $\{p_{(\alpha,U)}\}_{(\alpha,U)}^u$ satisfies

(a) For each $(\alpha, U)$, $p_{(\alpha,U)}$ is a real-analytic function of $u \in [0, \varepsilon)$;
(b) For $u = 0$, there is no mutation, i.e. $p_{(\alpha,U)} = 0$ for all $U \neq \emptyset$;
(c) For $u > 0$, the Fixation Axiom with Mutation is satisfied;
(d) The marginal probabilities $p_\alpha = \sum_{U \subseteq G} p_{(\alpha,U)}$ are independent of $u$;
(e) For each $n \geq 1$, $\mathbb{P}_{(\alpha,U)}[|U| \geq n] = \mathcal{O}(u^n)$.

Part (e) introduces a "big O" notation for asymptotic expansions in $u$:

$$f(u) = \mathcal{O}(u^n) \iff \lim_{u \to 0} \frac{f(u)}{u^n} < \infty. \quad (5.1)$$

We take these assumptions to be in force going forward. We also assume, for simplicity, that there is no self-mutation: $v_{a \to a} = 0$ for all $a \in A$. We denote the neutral drift and coalescent processes arising from this family of transition rules by $\widetilde{\mathcal{M}}_u$ and $\widetilde{C}_u$, respectively.

### 5.2. Mutant-appearance distribution

The stationary distribution, $\pi_{\widetilde{\mathcal{M}}_u}$, has a well-defined limit as $u \to 0$. In this limit, only monoallelic states are present, with probabilities given by the stationary distribution $\pi_A$ on $A$:

$$\lim_{u \to 0} \pi_{\widetilde{\mathcal{M}}_u}(\mathbf{x}) = \begin{cases} \pi_A(a) & \mathbf{x} = \mathbf{m}^a \text{ for some } a \in A, \\ 0 & \text{otherwise.} \end{cases} \quad (5.2)$$

This follows from Theorem 2 of Fudenberg and Imhof (2006), and can also be proven using Corollary 4.1.

To quantify genetic assortment under low mutation, it is necessary to characterize the states that arise when a mutation appears in a monoallelic state. This motivates the *mutant-appearance distribution* (Allen and Tarnita, 2014; Allen and McAvoy, 2019):

**Definition.** The *mutant-appearance distribution*, $\mu$, is the probability distribution on $A^G$ given by

$$\mu(\mathbf{x}) := \lim_{u \to 0}\left(\sum_{\mathbf{y} \in A^G} \pi_{\widetilde{\mathcal{M}}_u}(\mathbf{y}) \frac{P_{\mathbf{y} \to \mathbf{x}}}{1 - P_{\mathbf{y} \to \mathbf{y}}}\right) = \sum_{a \in A} \pi_A(a) \lim_{u \to 0}\left(\frac{P_{\mathbf{m}^a \to \mathbf{x}}}{1 - P_{\mathbf{m}^a \to \mathbf{m}^a}}\right). \quad (5.3)$$

In words, $\mu(\mathbf{x})$ is the limiting probability, as $u \to 0$, of state $\mathbf{x}$ arising when state $\mathbf{y}$ is sampled from $\lim_{u \to 0} \pi_{\widetilde{\mathcal{M}}_u}$, and then a transition occurs that leaves $\mathbf{y}$. The second equality follows from Eq. (5.2). The mutant appearance distribution is a natural distribution of initial states for Markov chains representing drift and/or selection.

There is a more explicit characterization of $\mu$ in terms of mutation probabilities. Let us denote $v'_S := \left.\frac{dv_S}{du}\right|_{u=0}$, where $v_S = \mathbb{E}_{(\alpha,U)}[|U \cap S|]$ is the expected number of mutations in set $S$ per transition, defined in Section 4.4. If, in a transition from state $\mathbf{m}^a$, a single new allele $a' \in A$ arises at site $g \in G$ by mutation, the resulting state has allele $a'$ in site $g$ and $a$ in all other sites. We denote this new state by $\mathbf{n}^{a,a'}_g \in A^G$. From Assumption (e) for low mutation, transition probabilities from a monoallelic state $\mathbf{m}^a$ can be expanded under low mutation as

$$P_{\mathbf{m}^a \to \mathbf{x}} = \begin{cases} 1 - u v'_G + \mathcal{O}(u^2) & \mathbf{x} = \mathbf{m}^a, \\ u v'_g v_{a \to a'} + \mathcal{O}(u^2) & \mathbf{x} = \mathbf{n}^{a,a'}_g \text{ for some } g \in G \text{ and } a, a' \in A, \\ \mathcal{O}(u^2) & \text{otherwise.} \end{cases} \quad (5.4)$$

Substituting into Eq. (5.3) yields the alternate formula:

$$\mu(\mathbf{x}) = \begin{cases} \dfrac{v'_g}{v'_G} \pi_A(a) v_{a \to a'} & \mathbf{x} = \mathbf{n}^{a,a'}_g \text{ for some } g \in G \text{ and } a, a' \in A, \\ 0 & \text{otherwise.} \end{cases} \quad (5.5)$$

The mutation-appearance distribution provides a link between the mutation-free chain $\mathcal{M}_0$ and the $u \to 0$ behavior of the chain with mutation, $\widetilde{\mathcal{M}}_u$. In particular, for any function $f : A^G \to \mathbb{R}$ that is zero on the monoallelic states we prove (Lemma E.3) that

$$\mathbb{E}_{\pi_{\widetilde{\mathcal{M}}_u}}[f] = u v'_G \sum_{t=0}^{\infty} \mathbb{E}_{\mathcal{M}_0}\left[f\left(\mathbf{X}^t\right) \mid \mathbf{X}^0 \sim \mu\right] + \mathcal{O}(u^2)$$

$$= u v'_G \sum_{t=0}^{\infty} \mathbb{E}_{C_0}\left[ f\left(\mathbf{X}_{C_t}\right) \mid \mathbf{X} \sim \mu \right] + \mathcal{O}(u^2). \tag{5.6}$$

With this result, one can pass back and forth between the transient behavior of $\mathcal{M}_0$ and the low-$u$ stationary behavior of $\widetilde{\mathcal{M}}_u$. A similar result was used by McAvoy and Allen (2021) to analyze fixation probabilities under weak selection.

### 5.3. Genetic assortment under low mutation

Our final main result shows that, under low mutation, statistics of identity-by-state, identity-by-descent, and mutations prior to coalescence, are directly related to each other, such that any can be obtained from any other.

**Theorem 5.1.** *For any nonempty set of sites $S \subseteq G$, the following quantities are equivalent to first order in $u$:*

- *The stationary probability, $\mathbb{E}_{\pi_{\widetilde{\mathcal{M}}_u}}\left[ \iota_S \right]$, that alleles in $S$ are identical-by-state,*
- *The probability, $q_S$, that alleles in $S$ are identical-by-descent,*
- *$1 - m_S$, where $m_S$ is the expected number of mutations prior to coalescence, and*
- *$1 - u v'_G \mathbb{E}_{\widetilde{\mathcal{M}}_0}\left[ T_S^{\mathrm{IBS}} \mid \mathbf{X}^0 \sim \mu \right]$, where $T_S^{\mathrm{IBS}}$ is the time that alleles in $S$ are not identical-by-state.*

*Moreover, the stationary probability that $S$ contains only allele $a \in A$ can be expanded as*

$$\mathbb{E}_{\pi_{\widetilde{\mathcal{M}}_u}}\left[ \iota_S^a \right] = \pi_{\mathcal{A}}(a)\, q_S + \mathcal{O}(u^2), \tag{5.7}$$

*and the stationary probability that $S$ contains two particular alleles, $a, a' \in A$, can be expanded as*

$$\mathbb{P}_{\pi_{\widetilde{\mathcal{M}}_u}}\left[ a, a' \in \{X_g\}_{g \in S} \right] = \left( \pi_{\mathcal{A}}(a)\, v_{a \to a'} + \pi_{\mathcal{A}}(a')\, v_{a' \to a} \right) m_S + \mathcal{O}(u^2). \tag{5.8}$$

The proof is in Appendix E. The practical utility of Theorem 5.1 arises from the fact that the first-order behavior of $m_S$ in $u$ can be computed systematically. Specifically, $m_S = u m'_S + \mathcal{O}(u^2)$, where the $m'_S$ are uniquely determined by the recurrence relation

$$m'_S = \begin{cases} v'_S + \sum_\alpha p_\alpha\, m'_{\alpha(S)} & |S| \geq 2, \\ 0 & |S| = 1, \end{cases} \tag{5.9}$$

which is the $u$-derivative of Eq. (4.3) at $u = 0$. We observe that $m'_S$ can be understood as coalescence length, rescaled by the local mutation rates $v'_g$. After solving Eq. (5.9), IBD probabilities can be obtained as $q_S = 1 - u m'_S + \mathcal{O}(u^2)$, stationary IBS probabilities as $\mathbb{E}_{\pi_{\widetilde{\mathcal{M}}_u}}\left[ \iota_S \right] = 1 - u m'_S + \mathcal{O}(u^2)$, and so on.

Theorem 5.1 builds on the relationship Slatkin (1991) observed between pairwise identity-by-descent and coalescence times. For low mutation, Theorem 5.1 extends this relationship to any set of sites, using the mutation-rescaled coalescence lengths, $m'_S$, as the relevant outcome of the coalescent process. Eq. (4.9) describes a similar relationship for larger mutation rates, but only in the case that mutation rates are constant over sites and independent of parentage.

## 6. Example: Diploid Wright–Fisher process with arbitrary sex ratio

We illustrate our framework and results by applying them to a diploid Wright–Fisher model with arbitrary sex ratio, complementing analyses by Nordborg and Krone (2002) and Wakeley (2009, §6.1–6.2). The case of even sex ratio was investigated by Möhle (1998a), and generalized beyond the Wright–Fisher model by Möhle and Sagitov (2003). Here, we obtain recurrence relations for fundamental quantities in the finite-population case, and then pass to the $N \to \infty$ limit to obtain explicit solutions and extend previous results.

### 6.1. Sites and parentage mappings

We consider a diploid two-sex population with discrete generations and random mating. Mutation occurs with probability $u \geq 0$ at each new allele, independently of all others.

Our first task is to describe how the genetic sites are partitioned among individuals and sexes. The individuals in the population are represented by a set $I$ of size $N$. The set $I$ is partitioned into subsets $I_{\mathrm{F}}$ and $I_{\mathrm{M}}$, of sizes $N_{\mathrm{F}}$ and $N_{\mathrm{M}}$, representing females and males, respectively. Each individual $i \in I$ contains two genetic sites, labeled $(i, 1)$ and $(i, 2)$, housing the allele inherited from the mother and father, respectively. The overall set of sites is therefore $G = \bigcup_{i \in I} G_i$, where $G_i := \{(i, 1), (i, 2)\}$ is the set of sites in individual $i$. $G$ can be partitioned into female and male sites, $G_{\mathrm{F}} := \bigcup_{i \in I_{\mathrm{F}}} G_i$, and $G_{\mathrm{M}} := \bigcup_{i \in I_{\mathrm{M}}} G_i$ respectively. $G$ can also be partitioned into maternally- and paternally-inherited sites, $G^1 := \{(i, 1) \mid i \in I\}$ and $G^2 := \{(i, 2) \mid i \in I\}$, respectively.

A parentage map $\alpha : G \to G$ with positive probability must satisfy $\alpha(G^1) \subseteq G_{\mathrm{F}}$ and $\alpha(G^2) \subseteq G_{\mathrm{M}}$ since all sites in $G^1$ are inherited from females and those in $G^2$ are inherited from males. All such maps are equally likely to be chosen. Explicitly, the probability of parentage map $\alpha : G \to G$ is

$$p_\alpha = \begin{cases} \left(\dfrac{1}{2 N_{\mathrm{F}}}\right)^N \left(\dfrac{1}{2 N_{\mathrm{M}}}\right)^N & \alpha(G^1) \subseteq G_{\mathrm{F}} \text{ and } \alpha(G^2) \subseteq G_{\mathrm{M}}, \\ 0 & \text{otherwise.} \end{cases} \tag{6.1}$$

This parentage distribution obeys the symmetry property that $p_{\alpha \circ \sigma} = p_\alpha$ for any $\alpha : G \to G$ and any permutation $\sigma$ of $G$ preserving $G^1$ and $G^2$.

### 6.2. Finite-population analysis

For this model, the recurrence relations for $\tau_S$, $\ell_S$, and $m_S$ – Eqs. (3.11), (3.12), and (4.3), respectively – all have the form

$$\psi_S = \begin{cases} f\left(\left|S \cap G^1\right|, \left|S \cap G^2\right|\right) + \sum_\alpha p_\alpha \psi_{\alpha(S)} & |S| \geq 2, \\ 0 & |S| = 1, \end{cases} \tag{6.2}$$

for particular functions $f(k^1, k^2)$, where $k^1 = |S \cap G^1|$ represents the number of maternally-inherited sites in $S$, and $k^2 = |S \cap G^2|$ the number of paternally-inherited sites. Lemma A.2 guarantees this system has a unique solution for $\{\psi_S\}_{S \subseteq G}$. By symmetry under permutations preserving $G^1$ and $G^2$, the solution $\psi_S$ for each $S$ depends only on $k^1 = |S \cap G^1|$ and $k^2 = |S \cap G^2|$. Eq. (6.2) therefore reduces to

$$\psi_{(k^1, k^2)} = \begin{cases} f(k^1, k^2) + \sum_{(\ell^1, \ell^2)} P_{(k^1, k^2) \to (\ell^1, \ell^2)}\, \psi_{(\ell^1, \ell^2)} & k^1 + k^2 \geq 2, \\ 0 & k^1 + k^2 = 1, \end{cases} \tag{6.3}$$

where

$$P_{(k^1, k^2) \to (\ell^1, \ell^2)} := \mathbb{P}_\alpha\left[ \left(\left|\alpha(S) \cap G^1\right|, \left|\alpha(S) \cap G^2\right|\right) = (\ell^1, \ell^2) \,\middle|\, \left(\left|S \cap G^1\right|, \left|S \cap G^2\right|\right) = (k^1, k^2) \right]. \tag{6.4}$$

We now derive an expression for $P_{(k^1, k^2) \to (\ell^1, \ell^2)}$. Consider an arbitrary nonempty set $S \subseteq G$, and let $k^1 = |S \cap G^1|$ and $k^2 = |S \cap G^2|$ as above. Under Eq. (6.1), each element of $S \cap G^1$ is mapped to a site chosen independently, uniformly at random, from $G_{\mathrm{F}}$. The probability of mapping $j$ sites to $G_{\mathrm{F}}^1 := G^1 \cap G_{\mathrm{F}}$ and the remaining $k^1 - j$ sites to $G_{\mathrm{F}}^2 := G^2 \cap G_{\mathrm{F}}$ is $\binom{k^1}{j} \left(\dfrac{1}{2}\right)^{k^1}$. Among the $j$ sites mapped independently to $G_{\mathrm{F}}^1$, the probability that they map to $\ell_{\mathrm{F}}^1$ *distinct* sites is the product of three factors: (i) the probability, $\left(\dfrac{1}{N_{\mathrm{F}}}\right)^j$, of any particular mapping of $j$ sites from $G_{\mathrm{F}}^1$; (ii) the number of ways, $\left\{\begin{matrix} j \\ \ell_{\mathrm{F}}^1 \end{matrix}\right\}$, of partitioning the $j$ sites into $\ell_{\mathrm{F}}^1$ nonempty subsets (a Stirling number of the second kind; Graham et al., 1994); and (iii) the number of ways, $\frac{N_F!}{(N_F-\ell_F^1)!}$, of assigning these $\ell_F^1$ nonempty subsets to distinct elements of $G_F^1$. So, among the $j$ elements of $S \cap G^1$ mapping to $G_F^1$, the probability there are $\ell_F^1$ distinct parent sites is $\left(\frac{1}{N_F}\right)^j \frac{N_F!}{(N_F-\ell_F^1)!} \left\{{j \atop \ell_F^1}\right\}$. By the same reasoning, among the $k^1 - j$ sites mapped to $G_F^2$, the probability of having $\ell_F^2$ distinct sites is $\left(\frac{1}{N_F}\right)^{k^1-j} \frac{N_F!}{(N_F-\ell_F^2)!} \left\{{k^1-j \atop \ell_F^2}\right\}$. Thus overall, the probability that there are $\ell_F^1$ distinct sites in $G_F^1$, and $\ell_F^2$ distinct sites in $G_F^2$, among the parents of sites in $S \cap G^1$ is

$$\mathbb{P}_F\left[(\ell_F^1, \ell_F^2) \mid k^1\right]$$
$$= \frac{1}{(2N_F)^{k^1}} \sum_{j=\ell_F^1}^{k^1-\ell_F^2} \binom{k^1}{j} \left(\frac{N_F!}{(N_F-\ell_F^1)!}\left\{{j \atop \ell_F^1}\right\}\right) \left(\frac{N_F!}{(N_F-\ell_F^2)!}\left\{{k^1-j \atop \ell_F^2}\right\}\right). \quad (6.5)$$

By a similar argument, the probability that there are $\ell_M^1$ distinct sites in $G_M^1 := G^1 \cap G_M$, and $\ell_M^2$ distinct sites in $G_M^2 := G^2 \cap G_M$ among the parents of sites in $S \cap G^2$, is

$$\mathbb{P}_M\left[(\ell_M^1, \ell_M^2) \mid k^2\right]$$
$$= \frac{1}{(2N_M)^{k^2}} \sum_{j=\ell_M^1}^{k^2-\ell_M^2} \binom{k^2}{j} \left(\frac{N_M!}{(N_M-\ell_M^1)!}\left\{{j \atop \ell_M^1}\right\}\right) \left(\frac{N_M!}{(N_M-\ell_M^2)!}\left\{{k^2-j \atop \ell_M^2}\right\}\right). \quad (6.6)$$

Putting everything together, we obtain an expression for the transition probability in Eq. (6.4):

$$P_{(k^1,k^2) \to (\ell^1, \ell^2)} = \sum_{\substack{(\ell_F^1, \ell_F^2), (\ell_M^1, \ell_M^2) \\ \ell_F^1 + \ell_M^1 = \ell^1, \ \ell_F^2 + \ell_M^2 = \ell^2}} \mathbb{P}_F[(\ell_F^1, \ell_F^2) \mid k^1] \, \mathbb{P}_M[(\ell_M^1, \ell_M^2) \mid k^2]. \quad (6.7)$$

It is helpful to write Eq. (6.3) in matrix form. For $1 \le \ell \le k \le N$, consider the $(k+1) \times (\ell+1)$ matrix $P^{(k,\ell)}$, whose $(k^1, \ell^1)$ entry, for $0 \le k^1 \le k$ and $0 \le \ell^1 \le \ell$, is $P_{(k^1,k-k^1) \to (\ell^1, \ell-\ell^1)}$. (Note that the rows and columns of $P^{(k,\ell)}$ are indexed starting at zero rather than one.) The entries of $P^{(k,\ell)}$ are the probabilities of all possible ways that $k$ sites can have $\ell$ distinct parental sites. Let $\psi^{(k)}$ and $f^{(k)}$ be $(k+1) \times 1$ vectors with $\psi^{(k)}_{k^1} = \psi_{(k^1,k-k^1)}$ and $f^{(k)}_{k^1} = f(k^1, k-k^1)$. Then, for $k^1 + k^2 \ge 2$, Eq. (6.3) can be rewritten in matrix form as

$$\psi^{(k)} = f^{(k)} + \sum_{\ell=2}^k P^{(k,\ell)} \psi^{(\ell)} = f^{(k)} + \sum_{\ell=2}^{k-1} P^{(k,\ell)} \psi^{(\ell)} + P^{(k,k)} \psi^{(k)}. \quad (6.8)$$

Since $P^{(k,k)}$ is substochastic, $I - P^{(k,k)}$ is invertible (Kemeny and Snell, 1976, Theorem 1.11.1). Solving Eq. (6.8) for $\psi^{(k)}$, we obtain

$$\psi^{(k)} = \begin{cases} \left(I - P^{(k,k)}\right)^{-1} f^{(k)} + \sum_{\ell=2}^{k-1} \left(I - P^{(k,k)}\right)^{-1} P^{(k,\ell)} \psi^{(\ell)} & k \ge 2, \\ \mathbf{0} & k = 1. \end{cases} \quad (6.9)$$

Eq. (6.9) allows for systematic computation of quantities arising from this coalescent process. As an example, for sets of size two, Eq. (6.3) becomes

$$\psi_{(0,2)} = f(0,2) + \frac{1}{4} \frac{N_M - 1}{N_M} \psi_{(0,2)} + \frac{1}{2} \psi_{(1,1)} + \frac{1}{4} \frac{N_M - 1}{N_M} \psi_{(2,0)}; \quad (6.10a)$$

$$\psi_{(1,1)} = f(1,1) + \frac{1}{4} \psi_{(0,2)} + \frac{1}{2} \psi_{(1,1)} + \frac{1}{4} \psi_{(2,0)}; \quad (6.10b)$$

$$\psi_{(2,0)} = f(2,0) + \frac{1}{4} \frac{N_F - 1}{N_F} \psi_{(0,2)} + \frac{1}{2} \psi_{(1,1)} + \frac{1}{4} \frac{N_F - 1}{N_F} \psi_{(2,0)}. \quad (6.10c)$$

The coefficients of $\psi_{(k^1,k^2)}$ on the right-hand side are the entries of $P^{(2,2)}$. For example, in Eq. (6.10a), $P_{(0,2) \to (0,2)} = \frac{1}{4} \frac{N_M - 1}{N_M}$ is the probability that two male-inherited alleles have different fathers ($\frac{N_M - 1}{N_M}$) and both come from male-inherited sites in these fathers ($\frac{1}{4}$); $P_{(0,2) \to (1,1)} = \frac{1}{2}$ is the probability that, for two male-inherited alleles, one comes from a male-inherited site and one comes from a female-inherited site (possibly in the same father); and $P_{(0,2) \to (2,0)} = \frac{1}{4} \frac{N_M - 1}{N_M}$ is the probability that two male-inherited alleles have different fathers ($\frac{N_M - 1}{N_M}$) and both come from female-inherited sites in these fathers ($\frac{1}{4}$). For expected coalescence times, $\tau$, $f(2,0) = f(1,1) = f(0,2) = 1$. Solving according to Eq. (6.9), we obtain the coalescence times for all sets of size two:

$$\tau_{\{g,h\}} = \begin{cases} \frac{2}{N} \left(4 N_F N_M + N_F - N_M\right) & g, h \in G^1, \\ \frac{2}{N} \left(4 N_F N_M + N_F + N_M\right) & g \in G^1, h \in G^2 \text{ or } g \in G^2, h \in G^1, \\ \frac{2}{N} \left(4 N_F N_M - N_F + N_M\right) & g, h \in G^2. \end{cases} \quad (6.11)$$

In particular, for finite $N$, expected coalescence times $\tau_S$ depend on the number of maternally- and paternally-inherited sites in $S$. However, if we take $N \to \infty$ with female sex ratio $\frac{N_F}{N}$ converging to $r \in (0,1)$, then $\lim_{N \to \infty} \frac{\tau_S}{N} = 8r(1-r)$ for all sets $S$ of size two. This suggests a simplification in the large-population limit, which we develop in the next subsection.

### 6.3. Large-population limit

We now consider the limiting behavior of solutions to Eq. (6.3) as $N \to \infty$ and $\frac{N_F}{N} \to r \in (0,1)$, stipulating that the function $f : (\mathbb{N} \cup \{0\})^2 \to \mathbb{R}$ does not vary with $N$.

Eqs. (6.5)–(6.7) imply that $P^{(k,\ell)}_{k^1,\ell^1}$ is of order $N^{-(k-\ell)}$, in the sense that $0 < \lim_{N \to \infty} N^{k-\ell} P^{(k,\ell)}_{k^1,\ell^1} < \infty$ for each $k \ge 2$, $1 \le \ell \le k$, $0 \le k^1 \le k$, and $0 \le \ell^1 \le \ell$. Dividing Eq. (6.9) by $N$ and taking $N \to \infty$, we obtain a recurrence equation of the form

$$\widetilde{\psi}^{(k)} = \begin{cases} F^{(k)} f^{(k)} + \Psi^{(k-1)} \widetilde{\psi}^{(k-1)} & k \ge 2, \\ \mathbf{0} & k = 1, \end{cases} \quad (6.12)$$

where

$$\widetilde{\psi}^{(k)} := \lim_{N \to \infty} N^{-1} \psi^{(k)}; \quad (6.13a)$$

$$F^{(k)} := \lim_{N \to \infty} N^{-1} \left(I - P^{(k,k)}\right)^{-1}; \quad (6.13b)$$

$$\Psi^{(k-1)} := \lim_{N \to \infty} \left(N^{-1} \left(I - P^{(k,k)}\right)^{-1}\right) \left(N P^{(k,k-1)}\right). \quad (6.13c)$$

In Appendix F, we show that the matrices $F^{(k)}$ and $\Psi^{(k-1)}$ have a simple form, in which all rows are identical and proportional to probabilities in the binomial distributions $\text{Binom}(k, 1/2)$ and $\text{Binom}(k-1, 1/2)$, respectively. Specifically,

$$\lim_{N \to \infty} P^{(k,k)}_{k^1,\ell^1} = \frac{1}{2^k} \binom{k}{\ell^1} \quad \text{for } 0 \le k_1 \le k \text{ and } 0 \le \ell_1 \le k; \quad (6.14a)$$

$$F^{(k)}_{k^1,\ell^1} = \frac{8r(1-r)}{2^k \binom{k}{2}} \binom{k}{\ell^1} \quad \text{for } 0 \le k_1 \le k \text{ and } 0 \le \ell_1 \le k; \quad (6.14b)$$

$$\Psi^{(k-1)}_{k^1,\ell^1} = \frac{1}{2^{k-1}} \binom{k-1}{\ell^1} \quad \text{for } 0 \le k_1 \le k \text{ and } 0 \le \ell_1 \le k-1. \quad (6.14c)$$

These results suggest that, in the $N \to \infty$ limit, the ancestors of alleles in $S$ are equally likely to be found in male-inherited or female-inherited sites, independently of each other, so that the total number in each is binomially distributed with probability $1/2$. Moreover, relaxation to this binomial distribution occurs on a faster timescale than the meeting of any subset of lineages.

In particular, since $F^{(k)}_{k^1,\ell^1}$ and $\Psi^{(k-1)}_{k^1,\ell^1}$ are independent of $k^1$, it follows from Eq. (6.12) that $\widetilde{\psi}_{(k^1,k^2)}$ depends only on $k = k^1 + k^2$ (that is, on the size of the set $S$). Accordingly, we replace $\widetilde{\psi}_{(k^1,k^2)}$ with $\widetilde{\psi}_k$. Let $\bar{f}(k)$ denote the expectation of $f(|S^1|, |S^2|)$ where $k$ elements are assigned independently to sets $S^1$ or $S^2$ with equal probability; explicitly,

$$\bar{f}(k) = \frac{1}{2^k} \sum_{k^1=0}^{k} \binom{k}{k^1} f(k^1, k - k^1). \tag{6.15}$$

Then, upon substituting from Eqs. (F.9) and (F.11), Eq. (6.12), for $k \geq 2$, reduces to

$$\widetilde{\psi}_k = \frac{8r(1-r)}{\binom{k}{2}} \bar{f}(k) + \widetilde{\psi}_{k-1}. \tag{6.16}$$

From this, we obtain the asymptotic solution to Eq. (6.3): for all $k^1, k^2$ with $k^1 + k^2 = k \geq 2$,

$$\lim_{N \to \infty} \frac{\Psi_{(k^1,k^2)}}{N} = \widetilde{\psi}_k = 8r(1-r) \sum_{j=2}^{k} \frac{\bar{f}(j)}{\binom{j}{2}}. \tag{6.17}$$

Applying this result, for $|S| \geq 2$, we have the following limits for coalescence times and lengths:

$$\lim_{N \to \infty} \frac{\tau_S}{N} = 16r(1-r)\left(1 - \frac{1}{|S|}\right); \tag{6.18a}$$

$$\lim_{N \to \infty} \frac{\ell_S}{N} = 16r(1-r) \sum_{k=1}^{|S|-1} \frac{1}{k}. \tag{6.18b}$$

Recalling that mutation occurs independently at each site with probability $u$, we have $\nu_S = u|S|$ for each subset $S$, and the expected number of mutations prior to coalescence in set $S$ is $m_S = u\ell_S$.

In the case of even sex ratio ($r = 1/2$) the factor $16r(1-r)$ reduces to 4. This recovers standard results for coalescence time and length in one-sex models (Hudson, 1990; Donnelly and Tavare, 1995), extended to two equal sexes by Möhle (1998a). For arbitrary sex ratio, $\tau_S$, $\ell_S$, and $m_S$ are rescaled by $4r(1-r)$ relative to the equal-sex case, in agreement with results from Nordborg and Krone (2002) and Wakeley (2009, §6.1–6.2).

These results can be used, via Theorem 5.1, to compute statistics of genetic assortment under low mutation. For any fixed, nonempty $S \subseteq G$, noting that $\nu'_G = N$, Theorem 5.1 gives

$$\begin{aligned}
\lim_{N \to \infty} \mathbb{E}_{\mathcal{M}_0}\left[T_S^{\text{IBS}} \mid \mathbf{X}^0 \sim \mu\right] &= \lim_{N \to \infty} \lim_{u \to 0} \frac{\mathbb{E}_{\pi_{\widetilde{\mathcal{M}}_u}}[1 - \iota_S]}{Nu} \\
&= \lim_{N \to \infty} \lim_{u \to 0} \frac{1 - q_S}{Nu} \\
&= \lim_{N \to \infty} \frac{m'_S}{N} \\
&= \lim_{N \to \infty} \frac{\ell_S}{N} \\
&= 16r(1-r) h_{|S|-1},
\end{aligned} \tag{6.19}$$

where $h_n = \sum_{k=1}^{n} \frac{1}{k}$ is the $n$th harmonic number. This means that, in the regime $N \gg 1$, $Nu \ll 1$, we have the approximations

$$\mathbb{E}_{\pi_{\widetilde{\mathcal{M}}_u}}[\iota_S] \approx q_S \approx 1 - 16Nur(1-r) h_{|S|-1}. \tag{6.20}$$

The errors in Eq. (6.20) are bounded by a function $\varepsilon(N, u)$ with $\lim_{N \to \infty} \lim_{u \to 0} \frac{\varepsilon(N,u)}{Nu} = 0$. We observe that the probability of a set $S$ being not IBS or IBD is $4r(1-r)$ times what it would be in an even-sex population. In similar fashion, given the Markov chain $\mathcal{A}$ for allele mutation, Theorem 5.1 can be combined with Eq. (6.18b), and $m'_S = \ell_S$, to approximate the stationary probability that $S$ contains a particular allele, or two particular alleles, in the $N \gg 1$, $Nu \ll 1$ regime.

### 6.4. Continuous-time limit

To connect our results to classical formulations of the coalescent, we observe from Eq. (6.4) that the sequence of pairs $(k_t^1, k_t^2)_{t=0}^{\infty}$ is a Markov chain on $(\mathbb{N} \cup \{0\})^2 \setminus \{(0,0)\}$. For a given initial set with $k_0^1$ maternally-inherited sites, and $k_0^2$ paternally inherited sites, the state of this chain represents the numbers of ancestor alleles in maternally- and paternally-inherited sites.

Our analysis implies that this Markov chain has a large-population, continuous-time limit. To obtain this limit, let us fix an initial state $(k_0^1, k_0^2)$, define the rescaled time variable $T = t/N$, and take $N \to \infty$ with $N_F/N \to r$. In light of the limits established in Eq. (6.14), Theorem 1 of Möhle (1998b) (or equivalently, the separation-of-timescales argument of Nordborg and Krone, 2002) implies that this chain converges in finite-dimensional distributions to a continuous-time process, with two levels. First, the total number of ancestors, $k_T = k_T^1 + k_T^2$, is characterized by a continuous-time Markov chain with transitions $k \to k-1$ occurring at rate $\binom{k}{2}/(8r(1-r))$. Second, for any given $k_T = k$, $k_T^1$ has conditional distribution Binom$(k, \frac{1}{2})$, and is conditionally independent of $k_{T'}^1$ for all $T' \neq T$ given the values of $k_T$ and $k_{T'}$. Informally, one can say that changes in the total number of ancestors follow Kingman's coalescent with time rescaled by $1/(4r(1-r))$. Moreover, at any given time in this rescaled process, each ancestor is equally likely to be at a maternally- or paternally-inherited site, independently of all other ancestors and times.

## 7. Discussion

### 7.1. Advantages of an abstract approach

Our work introduces a coalescent process applicable to a broad class of finite, structured population models, unified under a common representation. Although the level of abstraction used here is not needed for most applications, its advantage is that fundamental properties and relationships can be immediately obtained as instances of general results, rather than re-derived in each new setting. We anticipate our formalism and results being useful for both detailed modeling of specific populations ("tactical" modeling, in a dichotomy proposed by May, 1973), and advancement of general theory ("strategic" modeling).

On the tactical side, the coalescent introduced here is flexible enough to accommodate many varieties of complex structure (spatial, group, class, age, etc.) found in real-world populations. The major requirement is that the population size and structure remain fixed over time. For any model satisfying our assumptions, statistics of coalescence time and length, identity-by-descent, and mutation numbers can be computed from Eqs. (3.11), (3.12), (4.3) and (4.7), respectively. These recurrence equations have a convenient form, in that they can be solved recursively for larger sets in terms of smaller ones. For low mutation, statistics of identity-by-state and identity-by-descent can be computed, to first order in the mutation rate, by solving Eq. (5.9) for $m'_S$ and applying Theorem 5.1.

On the strategic side, our approach allows for the development of general theory regarding genetic assortment and its consequences for evolution. Coalescent duality (Theorem 3.1) and the construction of the stationary distribution (Section 4.5) provide tools to characterize the population states that arise under neutral drift. These states, in turn, form a baseline for characterizing natural selection, especially on alleles that only weakly affect fitness (Wild and Traulsen, 2007; Tarnita and Taylor, 2014; McAvoy and Allen, 2021). Our results may therefore be useful in further developing the mathematical theory of natural selection in structured populations, which is continuing to advance rapidly (Kirkpatrick et al., 2002; Rousset, 2004; Van Cleve, 2015; Lehmann et al., 2016; Allen and McAvoy, 2019; Mullon et al., 2021).

### 7.2. Application to social evolution

A particularly promising application of this work is to the evolution of social behavior—meaning behavior that affects the fitness of others.

Selection for social behavior depends on the co-occurrence of genes among those performing a behavior and those affected by it (Hamilton, 1964; Michod, 1982; Rousset, 2004; Nowak et al., 2010). For simple interactions, in which individuals make additive contributions (positive or negative) to fitness, the outcome of selection depends only on pairwise statistics of assortment (Hamilton, 1964; Michod and Hamilton, 1980; Taylor et al., 2007; Allen et al., 2017). Similarly, if mutations have only a small effect on phenotype, then the direction of selection depends only on pairwise genetic assortment (Taylor and Frank, 1996; Rousset and Billiard, 2000; Wild and Traulsen, 2007; Lehmann and Rousset, 2014; Van Cleve, 2015), while stability of evolutionary equilibria depends on triplet associations (Wakano and Lehmann, 2014; Mullon et al., 2016). However, for nonlinear, multilateral interactions, selection may depend on genetic associations among any number of individuals (Queller, 1985; Kirkpatrick et al., 2002; Gokhale and Traulsen, 2011; Ohtsuki, 2014; Van Cleve, 2015; Archetti, 2018). For small differences in fitness (but not necessarily in phenotype; see Wild and Traulsen, 2007), these associations may be computed at neutrality to predict the outcome of selection (Tarnita and Taylor, 2014; McAvoy and Allen, 2021; McAvoy and Wakeley, 2022).

In this way, our results can be applied to the analysis of nonlinear social evolution under weak selection. Under low mutation, Theorem 5.1 and Eq. (5.9) can be used to compute statistics of genetic association among any number of individuals, allowing for analysis of complex nonlinear interactions (Allen et al., 2024; Sheng et al., 2024). Similarly, Theorem 5.1 and Eq. (5.9) can be applied to compute the triplet associations needed to predict evolutionary stability and branching for quantitative traits that evolve by incremental mutation (Wakano and Lehmann, 2014; Mullon et al., 2016).

### 7.3. Extensions

#### 7.3.1. Large-population limit

By representing the entire ancestry of a finite-sized population, our formalism departs from the traditional focus of coalescent theory on finite samples from an infinite population. For simple population models, we are able to recover classical results by fixing attention on a subset of fixed size while taking the overall population size to infinity. However, it is not obvious how to achieve this in models with arbitrary structure. One approach would be to consider sequences of sets $\{G_j\}_{j\geq 1}$, and corresponding probability distributions $\{p_{(\alpha,U)}\}_{(\alpha,U)}^{j\geq 1}$, with $|G_j| \to \infty$, and identify conditions under which a well-defined $j \to \infty$ limit exists. Another approach would be to formulate a set-mapping-valued coalescent on an infinite set $G$, equipped with some probability distribution on the set of mappings $\alpha : G \to G$. For either route, there appear to be nontrivial mathematical challenges to overcome.

#### 7.3.2. Multilocus genetics

Additionally, multilocus extensions of the classical coalescent have been developed in a number of works (Hudson, 1983; Hudson and Kaplan, 1988; Barton and Navarro, 2002). To extend our framework to multilocus genetics, the set of sites $G$ must be expanded to include one site for each locus of interest, on each chromosome, within each individual. For example, $2Nk$ sites are needed to model $k$ loci in a diploid population of size $N$. Probabilities of recombination between loci can be incorporated into the parentage distribution, $\{p_\alpha\}_\alpha$. An alternative approach is to formulate a version of the ancestral recombination graph (Griffiths, 1991; Griffiths and Marjoram, 1996; Birkner et al., 2013) within the framework used here.

#### 7.3.3. Dynamic population structures

An assumption underlying our coalescent is that the population size and structure (spatial, mating pattern, age, etc.) is fixed. This limits its applicability to growing (Innan and Stephan, 2000) or fluctuating (Jagers and Sagitov, 2004; Pollak, 2010) populations, as well as populations for which the demographic (Tuljapurkar, 2013) or spatial structure (Lambert and Ma, 2015) varies with time.

For dynamic spatial structures, a path to generalization is provided by Su et al. (2023). They extend the formalism of parentage mappings to dynamic spatial structures, by including the current spatial structure as an aspect of the population state. Using this, they derive recurrence relations – analogous to our Eq. (3.11) – for coalescence times between pairs of sites. These are used to compute statistics of pairwise assortment, which are used in turn to derive conditions for selection in pairwise social behavior on a dynamic spatial network. Following this approach, the full coalescent introduced here may be generalizable to dynamic spatial structures. This would allow for computation of higher-order genetic associations, and, by extension, of conditions for (weak) selection with nonlinear, multilateral interactions in dynamic spatial environments.

Extending to populations of changing size would require more significant alteration to the formalism. One possibility is to allow for empty sites within the set $G$. Another is to let $G$ itself vary with the state. Either route requires care in constructing a coalescent with the appropriate duality properties. Similar challenges apply to models with variable numbers of individuals per class (Tuljapurkar, 2013) or per spatial location (e.g., Shpak et al., 2010), as well as models with continuous space (Barton et al., 2010).

### 7.4. Random mappings

Mathematically, this work expands on the connection between coalescent theory and the theory of random mappings (Katz, 1955; Harris, 1960; Kupka, 1990; Donnelly et al., 1991). This connection has been explored in the case of exchangeable random mappings (Pitman, 1999; Zubkov and Serov, 2015), corresponding to well-mixed populations. The more general class of random mappings satisfying the Fixation Axiom – representing neutral drift in structured populations – may be of both mathematical and biological interest.

**CRediT authorship contribution statement**

**Declaration of competing interest**

The authors declare that they have no known competing financial interests or personal relationships that could have appeared to influence the work reported in this paper.

**Data availability**

No data was used for the research described in the article.

**Acknowledgments**

The authors are grateful to John Wakeley and an anonymous reviewer for helpful feedback. This project was supported by Grant #62220 from the John Templeton Foundation, USA. Opinions expressed by the authors do not necessarily reflect the views of the funding agencies.

## Appendix A. Results for arbitrary finite Markov chains

This first appendix collects results pertaining to an arbitrary Markov chain $\mathcal{X} = (X^t)_{t=0}^{\infty}$ on a finite set $S$. For $x, y \in S$, let $P_{x \to y}$ denote the transition probability from $x$ to $y$, and let $P_{x \to y}^{(n)}$ denote the $n$-step transition probability from $x$ to $y$, for $n \geq 0$.

### A.1. Uniqueness of and convergence to stationary distribution

Our first result guarantees the existence of a unique stationary distribution – and convergence to that stationary distribution – for Markov chains that have some transient states but otherwise behave like regular chains.

**Lemma A.1.** *Let $R \subseteq S$ be a subset of the states of $\mathcal{X}$. Suppose that, for each $x \in R$, there is an $m \geq 0$ such that $P_{y \to x}^{(m)} > 0$ for each $y \in S$. Then all states in $R$ are positive recurrent, and all states not in $R$ are transient. Moreover, $\mathcal{X}$ has a unique stationary distribution $\pi_{\mathcal{X}}$, with $\pi_{\mathcal{X}}(y) = 0$ for each $y \notin R$, such that for all $x, y \in S$, $\lim_{n \to \infty} P_{x \to y}^{(n)} = \pi_{\mathcal{X}}(y)$.*

**Proof.** It is immediate that $R$ is the unique closed communicating class of $\mathcal{X}$, and all elements of $R$ are positive recurrent. By Theorem 6.2.1 of Kemeny and Snell (1976), there is a unique stationary distribution $\pi_{\mathcal{X}}$ on $\mathcal{X}$, with $\pi_{\mathcal{X}}(j) = 0$ for $j \notin R$.

To show that $\mathcal{X}$ converges to $\pi_{\mathcal{X}}$ from any initial state, we must show that $\mathcal{X}$ is aperiodic. For this, consider an arbitrary $x \in R$, and corresponding $m \geq 0$ satisfying the property given in the statement of the lemma. Then $P_{x \to x}^{(m)} > 0$. Moreover,

$$P_{x \to x}^{(m+1)} = \sum_{y \in S} P_{x \to y} P_{y \to x}^{(m)} > 0, \quad\quad (A.1)$$

since $P_{y \to x}^{(m)} > 0$ for each $y \in S$. Since transition from $x$ to itself is possible in both $m$ steps and $m+1$ steps, $x$ is an aperiodic state. Since $x$ is arbitrary, all elements of $R$ – that is, all recurrent states – are aperiodic. □

### A.2. Aggregating over transient behavior

We now consider the aggregation of a given function over the transient behavior of a Markov chain, from a given initial state. The following lemma shows that the moments of such aggregated quantities exist and are uniquely determined by a system of recurrence relations. This result will be used in Theorem C.1 to obtain recurrence relations for coalescent quantities.

**Lemma A.2.** *Let $S_T \subseteq S$ and $S_R \subseteq S$ denote the sets of transient and recurrent states of $\mathcal{X}$, respectively. Let $f : S \to \mathbb{R}$ be a function with $f(x) = 0$ for each $x \in S_R$. Let $F := \sum_{t=0}^{\infty} f(X^t)$. Then, for each $x \in S$ and $k \geq 0$, $\mathbb{E}_{\mathcal{X}}[F^k \mid X^0 = x]$ converges absolutely. Moreover, the recurrence relations*

$$\varphi_x^{(k)} = \begin{cases} \sum_{i=0}^{k} \binom{k}{i} (f(x))^{k-i} \sum_{y \in S} P_{x \to y} \varphi_y^{(i)} & x \in S_T, k \geq 1, \\ 0 & x \in S_R, k \geq 1, \\ 1 & k = 0 \end{cases} \quad (A.2)$$

*have unique solution $\varphi_x^{(k)} = \mathbb{E}_{\mathcal{X}}[F^k \mid X^0 = x]$ for each $x \in S$ and $k \geq 0$.*

**Proof.** Let $T_x^{S_R}$ denote the hitting time to $S_R$ from $x$ (i.e. the time to reach a recurrent state from initial state $x$). Given $X^0 = x$, we then have the bound $|F| \leq (\max_{y \in S} |f(y)|) T_x^{S_R}$. Since the time spent in transient states has an geometric tail (Kemeny and Snell, 1976, Corollary 3.1.2), all moments of $T_x^{S_R}$, and hence of $F$, converge for all $x \in S$.

To show that the $k$th moments, $\mathbb{E}_{\mathcal{X}}[F^k \mid X^0 = x]$, satisfy Eq. (A.2), we apply the binomial theorem:

$$\mathbb{E}_{\mathcal{X}}[F^k \mid X^0 = x] = \mathbb{E}_{\mathcal{X}}\left[\left(\sum_{t=0}^{\infty} f(X^t)\right)^k \mid X^0 = x\right]$$

$$= \sum_{i=0}^{k} \binom{k}{i} (f(x))^{k-i} \mathbb{E}_{\mathcal{X}}\left[\left(\sum_{t=1}^{\infty} f(X^t)\right)^i \mid X^0 = x\right]$$

$$= \sum_{i=0}^{k} \binom{k}{i} (f(x))^{k-i} \sum_{y \in S} P_{x \to y} \mathbb{E}_{\mathcal{X}}\left[\left(\sum_{t=0}^{\infty} f(X^t)\right)^i \mid X^0 = y\right]$$

$$= \sum_{i=0}^{k} \binom{k}{i} (f(x))^{k-i} \sum_{y \in S} P_{x \to y} \mathbb{E}_{\mathcal{X}}[F^i \mid X^0 = y]. \quad (A.3)$$

We prove uniqueness of the solution to Eq. (A.2) by strong induction on $k$. As a base case, suppose $k = 1$, so that Eq. (A.2) becomes

$$\varphi_x^{(1)} = \begin{cases} f(x) + \sum_{y \in S} P_{x \to y} \varphi_y^{(1)} & x \in S_T, \\ 0 & x \in S_R. \end{cases} \quad (A.4)$$

For any $T \geq 0$, iterating the $x \in S_T$ case of Eq. (A.4) gives

$$\varphi_x^{(1)} = \sum_{t=0}^{T} \mathbb{E}_{\mathcal{X}}[f(X^t) \mid X^0 = x] + \mathbb{E}_{\mathcal{X}}[\varphi_{X^{T+1}}^{(1)} \mid X^0 = x]. \quad (A.5)$$

Thus, any solution to Eq. (A.2) for $k = 1$ must satisfy Eq. (A.5) for all $T \geq 0$. As $T \to \infty$, the first term converges to $\mathbb{E}_{\mathcal{X}}[F \mid X^0 = x]$, while the second term, being absolutely bounded by $(\max_{y \in S} |f(y)|) \mathbb{P}[T_x^{\text{rec}} > T]$, converges to zero. This proves uniqueness of the solution to Eq. (A.2) in the base ($k = 1$) case.

For the induction step, consider $k \geq 2$ and suppose that, for all $1 \leq i \leq k-1$, the unique solution to Eq. (A.2) is $\varphi_x^{(i)} = \mathbb{E}_{\mathcal{X}}[F^i \mid X^0 = x]$ for all $x \in S$. Note that Eq. (A.2) can be written as

$$\varphi_x^{(k)} = \begin{cases} g(x) + \sum_{y \in S} P_{x \to y} \varphi_y^{(k)} & x \in S_T, \\ 0 & x \in S_R, \end{cases} \quad (A.6)$$

where, according to our inductive hypothesis,

$$g(x) = \sum_{i=0}^{k-1} \binom{k}{i} f(x)^{k-i} \sum_{y \in S} P_{x \to y} \mathbb{E}_{\mathcal{X}}[F^i \mid X^0 = y]. \quad (A.7)$$

By the argument used in the base case, the unique solution is

$$\varphi_x^{(k)} = \mathbb{E}_{\mathcal{X}}\left[\sum_{t=0}^{\infty} g(X^t) \mid X^0 = x\right], \quad (A.8)$$

which is equal to $\mathbb{E}_{\mathcal{X}}[F^k \mid X^0 = x]$ by Eq. (A.3). □

### A.3. Sojourn times and Markov chain perturbations

Again, let $S_T$ denote the set of transient states of $\mathcal{X}$, which we assume to be nonempty. For $x, y \in S_T$, we define the *sojourn time in $x$ from $y$* as

$$\sigma_y(x) := \sum_{t=0}^{\infty} \mathbb{P}_{\mathcal{X}}[X^t = x \mid X^0 = y]. \quad (A.9)$$

More generally, for any probability distribution $\beta$ on $S_T$, we define

$$\sigma_\beta(x) := \sum_{y \in S_T} \beta(y) \sigma_y(x) = \sum_{t=0}^{\infty} \mathbb{P}_{\mathcal{X}}[X^t = x \mid X^0 \sim \beta]. \quad (A.10)$$

There are two different recurrence relations for sojourn times. First, for fixed $x$, a recurrence for $\sigma_y(x)$ is obtained by applying Lemma A.2 (with $k = 1$) to the Kronecker delta function $f(z) = \delta_{z,x}$:

$$\sigma_y(x) = \delta_{y,x} + \sum_{z \in S_T} P_{y \to z} \sigma_z(x). \quad (A.11)$$

Second, for fixed $\beta$, we have

$$\sigma_\beta(x) = \beta(x) + \sum_{t=1}^{\infty} \mathbb{P}_{\mathcal{X}}\left[X^t = x \mid X^0 \sim \beta\right]$$

$$= \beta(x) + \sum_{y \in S_T} \sum_{t=1}^{\infty} \mathbb{P}_{\mathcal{X}}\left[X^{t-1} = y \mid X^0 \sim \beta\right] P_{y \to x}$$

$$= \beta(x) + \sum_{y \in S_T} \sigma_\beta(y) P_{y \to x}. \tag{A.12}$$

That this recurrence relation uniquely determines $\sigma_\beta$ follows from reasoning similar to that used in the proof of Lemma A.2. Specifically, for any finite time $T \geq 0$, iterating Eq. (A.12) gives

$$\sigma_\beta(x) = \sum_{y \in S_T} \beta(y) \sum_{t=0}^{T} P_{y \to x}^{(t)} + \sum_{y \in S_T} \sigma_\beta(y) P_{y \to x}^{(T+1)}$$

$$= \sum_{t=0}^{T} \mathbb{P}_{\mathcal{X}}\left[X^t = x \mid X^0 \sim \beta\right] + \sum_{y \in S_T} \sigma_\beta(y) P_{y \to x}^{(T+1)}. \tag{A.13}$$

As $T \to \infty$, the second term vanishes since $x$ is transient; thus, the unique solution to Eq. (A.12) is the sojourn time given in Eq. (A.10).

We now consider a Markov chain that is perturbed, changing its behavior from absorbing to regular. The following theorem shows that, as the perturbation parameter goes to zero, the stationary probability of a recurrent state $x$ becomes proportional to the sojourn time in $x$ from a particular initial distribution. This will be used to obtain Eq. (5.6), which in turn is required for Theorem 5.1.

**Theorem A.3.** *Consider a family of Markov chains $\{\mathcal{X}_u\}_{u \in [0,\varepsilon)}$ on $S$, with transition probabilities $P_{x \to y}(u)$. Suppose that*

  (a) *The transition probabilities $P_{x \to y}(u)$ are twice-differentiable functions of $u$,*
  (b) *For $u > 0$, $\mathcal{X}_u$ is regular, with (unique) stationary distribution $\pi_{\mathcal{X}_u}$,*
  (c) *For $u = 0$, there is a nonempty set $S_T$ of transient states.*

*Let $S_R = S \setminus S_T$ be the set of recurrent states of $\mathcal{X}_0$. For $x \in S_T$, define*

$$p(x) := \sum_{y \in S_R} \left(\lim_{u \to 0} \pi_{\mathcal{X}_u}(y)\right) \left.\frac{d P_{y \to x}}{du}\right|_{u=0}. \tag{A.14}$$

*Define the probability distribution $\beta$ on $S_T$ by*

$$\beta(x) = \frac{p(x)}{\sum_{y \in S_T} p(y)}. \tag{A.15}$$

*Then, for each $x \in S_T$,*

$$\pi_{\mathcal{X}_u}(x) = u \sigma_\beta(x) \sum_{y \in S_T} p(y) + \mathcal{O}(u^2). \tag{A.16}$$

**Proof.** For $x \in S$, we denote the limiting stationary probability of $x$ by $\pi_{\mathcal{X}_0}(x) := \lim_{u \to 0} \pi_{\mathcal{X}_u}(x)$. We observe that $\pi_{\mathcal{X}_0}$ comprises a stationary distribution for $\mathcal{X}_0$. Since transient states have stationary probability zero, $\pi_{\mathcal{X}_0}(x) = 0$ for each $x \in S_T$. This gives the zeroth-order term in Eq. (A.16).

We now turn to the first-order term. For each $x \in S_T$ and $u > 0$, stationarity of $\pi_{\mathcal{X}_u}$ implies

$$\pi_{\mathcal{X}_u}(x) = \sum_{y \in S_R} \pi_{\mathcal{X}_u}(y) P_{y \to x}(u) + \sum_{z \in S_T} \pi_{\mathcal{X}_u}(z) P_{z \to x}(u). \tag{A.17}$$

We take the $u$-derivative at $u = 0$ (denoted by primes, $'$) observing that $\pi_{\mathcal{X}_0}(z) = 0$ and $P_{y \to x}(0) = 0$ for all $x, z \in S_T$ and $y \in S_R$:

$$\pi'_{\mathcal{X}_u}(x) = \sum_{y \in S_R} \pi_{\mathcal{X}_0}(y) P'_{y \to x} + \sum_{z \in S_T} \pi'_{\mathcal{X}_u}(z) P_{z \to x}(0)$$

$$= p(x) + \sum_{z \in S_T} \pi'_{\mathcal{X}_u}(z) P_{z \to x}(0). \tag{A.18}$$

Dividing both sides by $\sum_{y \in S_T} p(y)$, we obtain

$$\frac{\pi'_{\mathcal{X}_u}(x)}{\sum_{y \in S_T} p(y)} = \beta(x) + \sum_{z \in S_T} \frac{\pi'_{\mathcal{X}_u}(z)}{\sum_{y \in S_T} p(y)} P_{z \to x}(0). \tag{A.19}$$

Comparing to Eq. (A.12) (the solution to which is unique), we have

$$\frac{\pi'_{\mathcal{X}_u}(x)}{\sum_{y \in S_T} p(y)} = \sigma_\beta(x), \tag{A.20}$$

for each $x \in S_T$. This provides the first-order term in Eq. (A.16). The result follows by Taylor's theorem (Spivak, 1994, p. 383). □

As a consequence, we obtain that stationary expectations of the regular chain, $\mathcal{X}_u$, are proportional, to first order in $u$, to transient sums over the absorbing chain, $\mathcal{X}_0$, from the initial distribution $\beta$:

**Corollary A.4.** *Under the assumptions of Theorem A.3, for any function $f : S \to \mathbb{R}$ with $f(x) = 0$ whenever $x$ is recurrent,*

$$\mathbb{E}_{\pi_{\mathcal{X}_u}}[f] = u \left(\sum_{x \in S_T} p(x)\right) \sum_{t=0}^{\infty} \mathbb{E}_{\mathcal{X}_0}\left[f(X^t) \mid X^0 \sim \beta\right] + \mathcal{O}(u^2), \tag{A.21}$$

*with $p(x)$ defined by Eq. (A.14) and $\beta$ defined by Eq. (A.15).*

**Proof.** Since the sum in the first-order term converges absolutely, we can write

$$\sum_{t=0}^{\infty} \mathbb{E}_{\mathcal{X}_0}\left[f(X^t) \mid X^0 \sim \beta\right] = \sum_{x \in S_T} \sigma_\beta(x) f(x). \tag{A.22}$$

Applying Theorem A.3 yields the desired result. □

**Appendix B. Stationary distribution of the coalescent**

Turning now to the coalescent process, we prove that $C$ has a unique stationary distribution, in which nonconstant mappings have probability zero, and constant mappings have probability proportional to the reproductive value of the corresponding site.

**Theorem B.1.** *There is a unique probability distribution $\{\rho_g\}_{g \in G}$ on $G$ such that*

$$\lim_{t \to \infty} \mathbb{P}\left[C_t = c\right] = \begin{cases} \rho_g & c = \kappa_g \text{ for some } g \in G, \\ 0 & \text{otherwise}. \end{cases} \tag{B.1}$$

*Moreover, the $\{\rho_g\}_{g \in G}$ are the unique solution to the system of equations*

$$\rho_g = \sum_\alpha p_\alpha \sum_{h \in \alpha^{-1}(g)} \rho_h; \tag{B.2a}$$

$$\sum_{g \in G} \rho_g = 1. \tag{B.2b}$$

**Proof.** Fix a site $g \in G$ satisfying the properties specified in the Fixation Axiom. This means there exists $m \geq 0$ and a sequence of replacement events $\{\alpha_k\}_{k=1}^{m}$, with each $p_{\alpha_k} > 0$, such that $\alpha_1 \circ \cdots \circ \alpha_m = \kappa_g$. Furthermore, for any set mapping $c : G \to G$, we have $\alpha_1 \circ \cdots \circ \alpha_m \circ c = \kappa_g$. Thus, for any state $c$ of $C$, there is a sequence of $m$ transitions, each with nonzero probability, leading to state $\kappa_g$. By Lemma A.1, $C$ converges to a unique stationary distribution, $\pi_C$, which is nonzero only for constant mappings $\kappa_g$, with $g$ satisfying the properties of the Fixation Axiom. Eq. (B.2) follows from the stationarity of $\pi_C$:

$$\rho_g = \pi_C(\kappa_g)$$

$$= \sum_{h \in G} \pi_C(\kappa_h) \sum_{\alpha : \alpha(h) = g} p_\alpha$$

$$= \sum_\alpha p_\alpha \sum_{h \in \alpha^{-1}(g)} \pi_C(\kappa_h)$$

$$= \sum_\alpha p_\alpha \sum_{h \in \alpha^{-1}(g)} \rho_h. \tag{B.3}$$

Uniqueness of the solution to (B.2) follows from uniqueness of the stationary distribution on $C$, as guaranteed by Lemma A.1. □

## Appendix C. Recurrence relations

We now use Lemma A.2 to obtain recurrence equations for the $k$th moments of coalescence time, $\tau_S^{(k)} := \mathbb{E}_C\left[\left(T_S^{\text{coal}}\right)^k\right]$; coalescence length, $\ell_S^{(k)} := \mathbb{E}_C\left[L_S^k\right]$; and mutations prior to coalescence, $m_S^{(k)} := \mathbb{E}_{\widetilde{C}}\left[M_S^k\right]$. Eqs. (3.11), (3.12), and (4.3) are obtained in the case $k = 1$.

**Theorem C.1.** *The $k$th moments of the coalescence times $T_S^{\text{coal}}$, coalescence lengths $L_S$, and mutation numbers $M_S$, are uniquely determined by the respective recurrence relations*

$$\tau_S^{(k)} = \begin{cases} 1 + \sum_{i=1}^{k} \binom{k}{i} \sum_{\alpha} p_{\alpha} \tau_{\alpha(S)}^{(i)} & |S| \geq 2, \\ 0 & |S| = 1; \end{cases} \quad \text{(C.1a)}$$

$$\ell_S^{(k)} = \begin{cases} |S|^k + \sum_{i=1}^{k} \binom{k}{i} |S|^{k-i} \sum_{\alpha} p_{\alpha} \ell_{\alpha(S)}^{(i)} & |S| \geq 2, \\ 0 & |S| = 1; \end{cases} \quad \text{(C.1b)}$$

$$m_S^{(k)} = \begin{cases} \sum_{(\alpha,U)} p_{(\alpha,U)} \sum_{i=0}^{k} \binom{k}{i} |U \cap S|^{k-i} m_{\alpha(S)}^{(i)} & |S| \geq 2, \\ 0 & |S| = 1. \end{cases} \quad \text{(C.1c)}$$

Above, we use the convention that the zeroth moment of any random variable is 1.

**Proof.** For Eqs. (C.1a)–(C.1b), we apply Lemma A.2 to the Markov chain $S$ defined in Section 3.2.1, noting that all subsets of size $|S| \geq 2$ are transient. Eq. (C.1a) follows from applying Lemma A.2 with $f(S) = 1$ when $|S| \geq 2$ and $f(S) = 0$ when $|S| = 1$. Similarly, Eq. (C.1b) follows from applying Lemma A.2 with $f(S) = |S|$ when $|S| \geq 2$ and $f(S) = 0$ when $|S| = 1$.

For Eq. (C.1c), we apply Lemma A.2 to the Markov chain $\widetilde{S} = (S_t, U_t)_{t=0}^{\infty}$ – constructed from $S$ in the same manner as $\widetilde{C}$ is from $C$ – using the function $f(S, U) = |U \cap S|$. □

## Appendix D. Stationary distribution on $\widetilde{\mathcal{M}}$

Here we establish that the neutral drift Markov chain with mutation, $\widetilde{\mathcal{M}}$, converges to a unique stationary distribution, which can be constructed by the procedure in Section 4.5.

**Lemma D.1.** *$\widetilde{\mathcal{M}}$ has a unique stationary distribution, $\pi_{\widetilde{\mathcal{M}}}$, with the property that $\lim_{t\to\infty} \mathbb{P}\left[\mathbf{X}^t = \mathbf{y}|\mathbf{X}^0 = \mathbf{x}\right] = \pi_{\widetilde{\mathcal{M}}}(\mathbf{y})$ for each $\mathbf{x}, \mathbf{y} \in A^G$.*

**Proof.** Let $A_0 \subseteq A$ be the unique closed communicating class of $\mathcal{A}$, and fix $a \in A_0$. We will show it is possible to reach state $\mathbf{m}^a$ from an arbitrary initial state $\mathbf{x}^0$.

Choose a site $g \in G$, with the properties specified by the Fixation Axiom with Mutation, and set $a_0 = x_g^0$. Since $a \in A_0$ and $A$ is finite, there exists a finite sequence $a_1, \ldots, a_k \in A$, with $a_k = a$, such that $v_{a_{i-1} \to a_i} > 0$ for each $i = 1, \ldots, k$. Since $\mathbb{P}_{(\alpha,U)}[g \in U] > 0$, it is possible (with positive probability) for the first $k$ transitions in $\widetilde{\mathcal{M}}$ to have $g \in U$, with mutation in $g$ from $a_{t-1}$ to $a_t$ occurring at each time $t = 1, \ldots, k$. This sequence results in $X_g^k = a_k = a$. Now suppose the next $k'$ transition events in $\widetilde{\mathcal{M}}$ are $(\alpha_1, \emptyset), \ldots, (\alpha_{k'}, \emptyset)$, with $\alpha_1 \circ \alpha_2 \circ \cdots \circ \alpha_{k'}(h) = g$ for all $h \in G$. (This sequence of events has positive probability by the assumed properties of $g$.) The resulting state is $\mathbf{X}^{k+k'} = \mathbf{m}^a$.

Thus $\mathbf{m}^a$ is accessible from an arbitrary initial state. Since the Fixation Axiom with Mutation implies $\mathbb{P}_{(\alpha,U)}[U = \emptyset] > 0$, transition from $\mathbf{m}^a$ to itself is possible. Therefore, there is some $m \geq 0$ such that $\mathbf{m}^a$ can be reached from any initial state in $m$ steps. The result then follows from Lemma A.1. □

**Theorem D.2.** *The random variable $\mathbf{X} \in A^G$ created by the construction in Section 4.5 has probability distribution $\pi_{\widetilde{\mathcal{M}}}$.*

**Proof.** We begin by defining a fixed-time variation on the construction in Section 4.5. First, sample $\widetilde{C}$ for time $T$. Next, sample $\mathbf{X}^0$ from $\pi_{\widetilde{\mathcal{M}}}$. Let $\mathbf{X}_{\text{anc}}^0 = (X_g^0)_{g \in C_T(G)}$ be the restriction of $\mathbf{X}^0$ to $C_T(G)$, representing the time-0 ancestors of the time-$T$ population. Last, iteratively construct a random sequence $(\mathbf{X}_{\text{anc}}^t)_{t=1}^T$, with each $\mathbf{X}_{\text{anc}}^t = (X_g^t)_{g \in C_{T-t}(G)} \in A^{C_{T-t}(G)}$ representing the time-$t$ ancestors of the time-$T$ population, by resolving mutations as in Step 3 of Section 4.5. Since this construction is compatible with the transition rule for $\widetilde{\mathcal{M}}$, and since $\mathbf{X}^0$ is sampled from the stationary distribution $\pi_{\widetilde{\mathcal{M}}}$, the final state, $\mathbf{X}_{\text{anc}}^T$ also has distribution $\pi_{\widetilde{\mathcal{M}}}$.

It remains to show that, instead of stopping $\widetilde{C}$ at a fixed time $T$ and sampling the initial state from $\pi_{\widetilde{\mathcal{M}}}$, we may stop at the time of coalescence and sample an initial allele from $\pi_{\mathcal{A}}$. For now, let us condition the fixed-time construction on $C_T$ being a constant map, so that there exists a coalescence time, $\hat{T} = T_G^{\text{coal}} \leq T$. Let $M = \sum_{t=\hat{T}}^{T} |U_t \cap C_t(G)|$ represent the number of mutations in the sequence $\mathbf{X}_{\text{anc}}^0, \ldots, \mathbf{X}_{\text{anc}}^{T-\hat{T}}$ (i.e. between the initial ancestor and the most recent common ancestor). Since $\mathbf{X}^0$ is sampled from $\pi_{\widetilde{\mathcal{M}}}$ independently of $\widetilde{C}$, $\mathbf{X}_{\text{anc}}^0 = X_{C_T(G)}^0$ has marginal distribution $\pi_{\mathcal{A}}$ by Eq. (4.1), independently of $M$. $\mathbf{X}_{\text{anc}}^{T-\hat{T}} = X_{C_{\hat{T}}(G)}^{T-\hat{T}}$ is then obtained by is obtained by taking $M$ steps in $\mathcal{A}$ starting from $\mathbf{X}_{\text{anc}}^0$. By stationarity and the independence of $M$ from $\mathbf{X}_{\text{anc}}^0$, $\mathbf{X}_{\text{anc}}^{T-\hat{T}}$ also has distribution $\pi_{\mathcal{A}}$.

To conclude, we set $\mathbf{Y}^t = \mathbf{X}_{\text{anc}}^{t+T-\hat{T}}$ for $0 \leq t \leq T - \hat{T}$. Then $\mathbf{Y}^{\hat{T}} = \mathbf{X}_{\text{anc}}^T$ has distribution $\pi_{\widetilde{\mathcal{M}}}$. Letting $T \to \infty$, so that the probability of $C_T$ being constant converges to one, we obtain a construction of $(\mathbf{Y}^t)_{t=1}^{\hat{T}}$ equivalent to that in Section 4.5. □

## Appendix E. Low-mutation asymptotics

Here we derive the low-mutation behavior of key quantities, under the assumptions of Section 5.1, leading up to the proof of Theorem 5.1. We begin with a technical lemma on the low-$u$ behavior of $M_S$:

**Lemma E.1.** *For every nonempty $S \subseteq G$ and $k \geq 0$, $\mathbb{P}_{\widetilde{C}_u}\left[M_S = k\right]$ is a real-analytic function of $u$ and is $\mathcal{O}(u^k)$. Moreover, for any function $h(j)$ of at most polynomial growth in $j$, $\sum_{j=k}^{\infty} h(j) \mathbb{P}_{\widetilde{C}_u}\left[M_S = j\right] = \mathcal{O}(u^k)$.*

**Proof.** We will prove $\mathbb{P}_{\widetilde{C}_u}\left[M_S = k\right]$ is $\mathcal{O}(u^k)$ and real-analytic by strong induction on $k$. For the base case ($k = 0$) we recall that $\mathbb{P}_{\widetilde{C}_u}\left[M_S = 0\right] = q_S$. For any $u \geq 0$, $q_S$ can be obtained from Eq. (4.7) via matrix inversion (Kemeny and Snell, 1976, Theorem 3.3.7). Since the $p_{(\alpha,U)}$ are real-analytic in $u$, and matrix inversion is real-analytic in the matrix entries, $q_S = \mathbb{P}_{\widetilde{C}_u}\left[M_S = 0\right]$ is real-analytic in $u$.

We now proceed to the induction step. Fix $k \geq 1$ and suppose that, for every nonempty $S \subseteq G$ and all $0 \leq j \leq k-1$, $\mathbb{P}_{\widetilde{C}_u}\left[M_S = j\right]$ is real-analytic and $\mathcal{O}(u^j)$ in $u$. For all sets $S$ with $|S| \geq 2$, we expand

$$\mathbb{P}_{\widetilde{C}_u}\left[M_S = k\right] = \sum_{j=0}^{k-1} \sum_{\substack{(\alpha,U) \\ |U \cap S| = k-j}} p_{(\alpha,U)} \mathbb{P}_{\widetilde{C}_u}\left[M_{\alpha(S)} = j\right]$$
$$+ \sum_{\substack{(\alpha,U) \\ U \cap S = \emptyset}} p_{(\alpha,U)} \mathbb{P}_{\widetilde{C}_u}\left[M_{\alpha(S)} = k\right]. \quad (\text{E.1})$$

Collecting the probabilities $\mathbb{P}_{\widetilde{C}_u}\left[M_S = k\right]$, for all nonempty $S \subseteq G$, into a vector $\mathbf{y}$, we can write Eq. (E.1) in vector-matrix form as $\mathbf{y} = \mathbf{b} + Q\mathbf{y}$. For all $u > 0$, the matrix $Q$ is substochastic since $\mathbb{P}_{(\alpha,U)}[U = \emptyset] < 1$. Therefore, $I - Q$ is invertible (Kemeny and Snell, 1976, Theorem 1.11.1) and $\mathbf{y} = (I - Q)^{-1}\mathbf{b}$. Real-analyticity of matrix inversion, together with the induction hypothesis, implies that $\mathbb{P}_{\widetilde{C}_u}\left[M_S = k\right]$ is real-analytic in $u$.

Dividing both sides of Eq. (E.1) by $u^k$ and taking $u \to 0$ yields:

$$\lim_{u\to 0} \frac{\mathbb{P}_{\widetilde{C}_u}[M_S = k]}{u^k}$$
$$= \sum_{j=0}^{k-1} \sum_{\substack{(\alpha,U) \\ |U \cap S| = k-j}} \left(\lim_{u\to 0} \frac{p_{(\alpha,U)}}{u^{k-j}}\right) \left(\lim_{u\to 0} \frac{\mathbb{P}_{\widetilde{C}_u}[M_{\alpha(S)} = j]}{u^j}\right)$$
$$+ \sum_{\substack{(\alpha,U) \\ U \cap S = \varnothing}} p_{(\alpha,U)} \left(\lim_{u\to 0} \frac{\mathbb{P}_{\widetilde{C}_u}[M_{\alpha(S)} = k]}{u^k}\right). \quad \text{(E.2)}$$

By the Low-Mutation Assumption and the induction hypothesis, the limits in the first term on the right-hand side all converge. Eq. (E.2) also has the form $\mathbf{y} = \mathbf{b} + Q\mathbf{y}$, with the same $Q$ as above but different $\mathbf{b}$. The same argument therefore guarantees a unique, finite solution for $\lim_{u\to 0} \frac{\mathbb{P}_{\widetilde{C}_u}[M_S = k]}{u^k}$ for each nonempty $S \subseteq G$. Thus, $\mathbb{P}_{\widetilde{C}_u}[M_S = k] = \mathcal{O}(u^k)$. By induction, $\mathbb{P}_{\widetilde{C}_u}[M_S = k]$ is real-analytic and $\mathcal{O}(u^k)$ for all $k$.

For the remaining claim, we note that $M_S \leq |S| T_S^{\text{coal}}$. Since $T_S^{\text{coal}}$ has a geometric tail and $h(j)$ has at most polynomial growth, we have, for each $k \geq 0$,

$$\sum_{j=k}^{\infty} h(j) \mathbb{P}_{\widetilde{C}_u}[M_S = j] \leq |S| \sum_{j=k}^{\infty} h(j) \mathbb{P}_{\widetilde{C}_u}[T_S^{\text{coal}} = j] < \infty. \quad \text{(E.3)}$$

By the Dominated Convergence Theorem, we may interchange sums and limits, leading to

$$\lim_{u\to 0} \frac{1}{u^k} \sum_{j=k}^{\infty} h(j) \mathbb{P}_{\widetilde{C}_u}[M_S = j] = \sum_{j=k}^{\infty} h(j) \lim_{u\to 0} \left(\frac{1}{u^k} \mathbb{P}_{\widetilde{C}_u}[M_S = j]\right)$$
$$= h(k) \lim_{u\to 0} \left(\frac{1}{u^k} \mathbb{P}_{\widetilde{C}_u}[M_S = k]\right) < \infty, \quad \text{(E.4)}$$

since $\mathbb{P}_{\widetilde{C}_u}[M_S = j] = \mathcal{O}(u^j)$ for all $j \geq k$. □

We next obtain a low-mutation expansion for the IBD probability, $q_S$, in terms of the moments of $M_S$. This expansion involves the signed Stirling numbers of the first kind, $s(n,k)$, which are defined by the equation (Graham et al., 1994):

$$x(x-1)\cdots(x-n+1) = \sum_{k=1}^{n} s(n,k) x^k. \quad \text{(E.5)}$$

**Proposition E.2.** *For each nonempty $S \subseteq G$, and each $n \geq 0$,*

$$q_S = \frac{(-1)^n}{n!} \sum_{k=0}^{n} s(n+1, k+1) m_S^{(k)} + \mathcal{O}(u^{n+1}). \quad \text{(E.6)}$$

This result provides a sequence of low-$u$ approximations of $q_S$ in terms of moments of $M_S$:

$$q_S = 1 - m_S + \mathcal{O}(u^2) \quad \text{(E.7a)}$$
$$= 1 - \frac{1}{2}\left(3m_S^{(1)} - m_S^{(2)}\right) + \mathcal{O}(u^3) \quad \text{(E.7b)}$$
$$= 1 - \frac{1}{6}\left(11 m_S^{(1)} - 6 m_S^{(2)} + m_S^{(3)}\right) + \mathcal{O}(u^4), \quad \text{(E.7c)}$$

and so on.

**Proof.** Consider the polynomial

$$p_n(x) := \frac{(-1)^n}{n!}(x-1)(x-2)\cdots(x-n)$$
$$= \frac{(-1)^n}{n!} \sum_{k=0}^{n} s(n+1, k+1) x^k. \quad \text{(E.8)}$$

For nonnegative integers $x$, the values of $p_n(x)$ are

$$p_n(x) = \begin{cases} 1 & x = 0, \\ 0 & 1 \leq x \leq n, \\ (-1)^n \binom{x-1}{n} & x \geq n+1. \end{cases} \quad \text{(E.9)}$$

We proceed by writing two expressions for the expectation of $p_n(M_S)$. First, taking expectations of Eq. (E.8), we obtain

$$\mathbb{E}_{\widetilde{C}_u}[p_n(M_S)] = \frac{(-1)^n}{n!} \sum_{k=0}^{n} s(n+1, k+1) m_S^{(k)}. \quad \text{(E.10)}$$

Second, applying Eq. (E.9),

$$\mathbb{E}_{\widetilde{C}_u}[p_n(M_S)] = \sum_{k=0}^{\infty} \mathbb{P}_{\widetilde{C}_u}[M_S = k] p_n(k)$$
$$= \mathbb{P}_{\widetilde{C}_u}[M_S = 0] + (-1)^n \sum_{k=n+1}^{\infty} \binom{k-1}{n} \mathbb{P}_{\widetilde{C}_u}[M_S = k]. \quad \text{(E.11)}$$

Setting the right-hand sides of Eqs. (E.10) and (E.11) equal, replacing $\mathbb{P}_{\widetilde{C}_u}[M_S = 0]$ by $q_S$, and rearranging, we obtain

$$q_S = \frac{(-1)^n}{n!} \sum_{k=0}^{n} s(n+1, k+1) m_S^{(k)}$$
$$+ (-1)^{n+1} \sum_{k=n}^{\infty} \binom{k}{n} \mathbb{P}_{\widetilde{C}_u}[M_S = k+1]. \quad \text{(E.12)}$$

Since $\binom{k}{n}$ grows polynomially in $k$ for fixed $n$, the second term on the right-hand side is $\mathcal{O}(u^{k+1})$ by Lemma E.1. □

We now prove Eq. (5.6), relating the transient behavior of $\widetilde{\mathcal{M}}_0$ to low-$u$ stationary behavior of $\widetilde{\mathcal{M}}_u$:

**Lemma E.3.** *If $f : A^G \to \mathbb{R}$ satisfies $f(\mathbf{m}^a) = 0$ for each $a \in A$, then*

$$\mathbb{E}_{\pi_{\widetilde{\mathcal{M}}_u}}[f] = u v_G' \sum_{t=0}^{\infty} \mathbb{E}_{\widetilde{\mathcal{M}}_0}\left[f(\mathbf{X}^t) \mid \mathbf{X}^0 \sim \mu\right] + \mathcal{O}(u^2)$$
$$= u v_G' \sum_{t=0}^{\infty} \mathbb{E}_{\widetilde{C}_0}\left[f(\mathbf{X}_{C_t}) \mid \mathbf{X} \sim \mu\right] + \mathcal{O}(u^2). \quad \text{(E.13)}$$

**Proof.** We assume $|G| \geq 2$, otherwise there are no non-monoallelic states and the statement is vacuous. Under this assumption, Eq. (5.5) implies that the mutant-appearance distribution is supported on non-monoallelic states. From Eq. (5.3), we have (using primes to indicate $u$-derivatives at $u = 0$)

$$\mu(\mathbf{x}) = \lim_{u\to 0} \left(\frac{\sum_{a\in A} \pi_{\mathcal{A}}(a) P_{\mathbf{m}^a \to \mathbf{x}}}{\mathbb{P}_{(\alpha,U)}[U \neq \varnothing]}\right)$$
$$= \frac{\sum_{a\in A} \pi_{\mathcal{A}}(a) P'_{\mathbf{m}^a \to \mathbf{x}}}{\mathbb{P}'_{(\alpha,U)}[U \neq \varnothing]}. \quad \text{(E.14)}$$

In particular, $\mu(\mathbf{x})$ is proportional to $\sum_{a\in A} \left(\lim_{u\to 0} \pi_{\widetilde{\mathcal{M}}_u}(\mathbf{m}^a)\right) P'_{\mathbf{m}^a \to \mathbf{x}}$. The first equality now follows from applying Corollary A.4, with $\mu$ playing the role of the probability distribution $\beta$. The second equality follows from coalescent duality, Theorem 3.1. □

Finally we prove our main result for low mutation, Theorem 5.1.

**Proof of Theorem 5.1.** Eq. (E.7a) gives $q_S = 1 - m_S + \mathcal{O}(u^2)$. From Lemma E.3, we have

$$\mathbb{E}_{\pi_{\widetilde{\mathcal{M}}_u}}[\iota_S] = 1 - \mathbb{E}_{\pi_{\widetilde{\mathcal{M}}_u}}[1 - \iota_S]$$
$$= 1 - u v_G' \sum_{t=0}^{\infty} \mathbb{E}_{\widetilde{\mathcal{M}}_0}\left[1 - \iota_S(\mathbf{X}^t) \mid \mathbf{X}^0 \sim \mu\right] + \mathcal{O}(u^2)$$
$$= 1 - u v_G' \mathbb{E}_{\widetilde{\mathcal{M}}_0}\left[T_S^{\text{IBS}} \mid \mathbf{X}^0 \sim \mu\right] + \mathcal{O}(u^2). \quad \text{(E.15)}$$

For all other equivalencies, we work with the pairing $(\widetilde{C}_u; \mathbf{X})$ defined in Section 4.5. For an arbitrary function $f : A^G \to \mathbb{R}$, we use Lemma E.1 and Proposition E.2 to expand

$$\begin{aligned}
\mathbb{E}_{\pi_{\widetilde{\mathcal{M}}_u}}[f] &= \mathbb{E}_{(\widetilde{c}_u;\mathbf{X})}[f] \\
&= \mathbb{E}_{(\widetilde{c}_u;\mathbf{X})}[f \mid M_S = 0] \, \mathbb{P}[M_S = 0] \\
&\quad + \mathbb{E}_{(\widetilde{c}_u;\mathbf{X})}[f \mid M_S = 1] \, \mathbb{P}[M_S = 1] + \mathcal{O}(u^2) \\
&= \mathbb{E}_{(\widetilde{c}_u;\mathbf{X})}[f \mid M_S = 0] \, q_S \\
&\quad + \mathbb{E}_{(\widetilde{c}_u;\mathbf{X})}[f \mid M_S = 1] \, (1 - q_S) + \mathcal{O}(u^2).
\end{aligned} \quad (\text{E.16})$$

The remaining equivalencies now follow from Corollary 4.1, using appropriate choices for $f$. Using $f = \iota_S$ gives $\mathbb{E}_{\pi_{\widetilde{\mathcal{M}}_u}}[\iota_S] = q_S + \mathcal{O}(u^2)$. Using $f = \iota_S^a$ gives Eq. (5.7). Finally, Eq. (5.8) is proven by letting $f$ be an indicator function that equals 1 if and only if $S$ contains alleles $a$ and $a'$. □

**Appendix F. Large-$N$ limits for diploid Wright-Fisher process**

Here we derive Eq. (6.14), which gives large-$N$ asymptotics of transition probabilities for the coalescent in the diploid Wright–Fisher process (Section 6).

We first compute $\lim_{N\to\infty} P^{(k,k)}$. For $\ell = k$, Eqs. (6.5) and (6.6) become

$$\mathbb{P}_{\mathrm{F}}\left[(\ell_{\mathrm{F}}^1, k^1 - \ell_{\mathrm{F}}^1) \mid k^1\right] = \frac{1}{(2N_{\mathrm{F}})^{k^1}} (k^1)! \binom{N_{\mathrm{F}}}{\ell_{\mathrm{F}}^1}\binom{N_{\mathrm{F}}}{k^1 - \ell_{\mathrm{F}}^1}; \quad (\text{F.1a})$$

$$\mathbb{P}_{\mathrm{M}}\left[(\ell_{\mathrm{M}}^1, k^2 - \ell_{\mathrm{M}}^1) \mid k^2\right] = \frac{1}{(2N_{\mathrm{M}})^{k^2}} (k^2)! \binom{N_{\mathrm{M}}}{\ell_{\mathrm{M}}^1}\binom{N_{\mathrm{M}}}{k^2 - \ell_{\mathrm{M}}^1}. \quad (\text{F.1b})$$

Taking the $N \to \infty$ limit of these expressions, summing their product over all $\ell_{\mathrm{F}}^1$ and $\ell_{\mathrm{M}}^1$ such that $\ell_{\mathrm{F}}^1 + \ell_{\mathrm{M}}^1 = \ell^1$, and using Vandermonde's identity, we obtain

$$\lim_{N\to\infty} P_{k^1,\ell^1}^{(k,k)} = \frac{1}{2^k} \sum_{\ell_{\mathrm{F}}^1 + \ell_{\mathrm{M}}^1 = \ell^1} \binom{k^1}{\ell_{\mathrm{F}}^1}\binom{k^2}{\ell_{\mathrm{M}}^1} = \frac{1}{2^k}\binom{k}{\ell^1}, \quad (\text{F.2})$$

verifying Eq. (6.14a). This shows that $\lim_{N\to\infty} P^{(k,k)}$ is a stochastic matrix with identical rows, each containing the probabilities of the binomial distribution $\mathrm{Binom}(k, 1/2)$. This means that, as $N \to \infty$, the number of ancestral lineages in female-inherited sites becomes binomially distributed with probability $1/2$, on a faster timescale than the meeting of any two lineages.

We next compute $F^{(k)} := \lim_{N\to\infty} N^{-1}\left(I - P^{(k,k)}\right)^{-1}$. Applying Jacobi's formula for the derivative of a determinant, we obtain

$$\begin{aligned}
F^{(k)} &= \lim_{N\to\infty} \frac{N^{-1}}{\det\left(I - P^{(k,k)}\right)} \operatorname{adj}\left(I - P^{(k,k)}\right) \\
&= \lim_{N\to\infty} \frac{-1}{\operatorname{tr}\left(\operatorname{adj}\left(I - P^{(k,k)}\right) N^2 \frac{d}{dN}\left(I - P^{(k,k)}\right)\right)} \operatorname{adj}\left(I - P^{(k,k)}\right) \\
&= \frac{1}{\operatorname{tr}\left(\left(\lim_{N\to\infty} P^{(k,k)}\right)\left(\lim_{N\to\infty} N^2 \frac{d}{dN} P^{(k,k)}\right)\right)} \left(\lim_{N\to\infty} P^{(k,k)}\right). \quad (\text{F.3})
\end{aligned}$$

The final step in Eq. (F.3) is justified by applying the following lemma to $\lim_{N\to\infty} P^{(k,k)}$:

**Lemma F.1.** *For any stochastic matrix $P$ whose rows are all equal, $\operatorname{adj}(I - P) = P$.*

**Proof.** Since $P$ is stochastic, it has eigenvalue one, therefore $\det(I - P) = 0$. By the properties of the adjugate matrix,

$$(I - P)\operatorname{adj}(I - P) = \operatorname{adj}(I - P)(I - P) = \det(I - P) I = 0, \quad (\text{F.4})$$

which implies that $P\operatorname{adj}(I - P) = \operatorname{adj}(I - P) P = \operatorname{adj}(I - P)$. Since the rows of $P$ are all equal, it follows that the rows of $\operatorname{adj}(I - P)$ are also all equal, and proportional to those of $P$, i.e. $\operatorname{adj}(I - P) = cP$ for some $c > 0$. Noting that $P^2 = P$ and hence $(I - P)^2 = I - P$, we have

$$cP = \operatorname{adj}(I - P) = \operatorname{adj}\left((I - P)^2\right) = (\operatorname{adj}(I - P))^2 = c^2 P^2 = c^2 P, \quad (\text{F.5})$$

and therefore $c = 1$. □

To continue with the computation of $F^{(k)}$, we evaluate $\lim_{N\to\infty} N^2 \frac{d}{dN} P^{(k,k)}$. For any polynomial of the form $p(N) = a_0 + a_1 N^{-1} + \cdots + a_k N^{-k}$, we observe that $\lim_{N\to\infty} N^2 \frac{d}{dN} p(N) = -a_1$. Using this, and recalling that $N_{\mathrm{F}}/N \to r$ as $N \to \infty$, we obtain

$$\lim_{N\to\infty} N^2 \frac{d}{dN} \mathbb{P}_{\mathrm{F}}\left[(\ell_{\mathrm{F}}^1, k^1 - \ell_{\mathrm{F}}^1) \mid k^1\right]$$
$$= r^{-1} \frac{1}{2^{k^1}} \binom{k^1}{\ell_{\mathrm{F}}^1}\left(\binom{\ell_{\mathrm{F}}^1}{2} + \binom{k^1 - \ell_{\mathrm{F}}^1}{2}\right); \quad (\text{F.6a})$$

$$\lim_{N\to\infty} N^2 \frac{d}{dN} \mathbb{P}_{\mathrm{M}}\left[(\ell_{\mathrm{M}}^1, k^2 - \ell_{\mathrm{M}}^1) \mid k^2\right]$$
$$= (1 - r)^{-1} \frac{1}{2^{k^2}} \binom{k^2}{\ell_{\mathrm{M}}^1}\left(\binom{\ell_{\mathrm{M}}^1}{2} + \binom{k^2 - \ell_{\mathrm{M}}^1}{2}\right). \quad (\text{F.6b})$$

Letting $k^2 = k - k^1$, the identity $\sum_{k=0}^{r} \binom{m}{k}\binom{n}{r-k}\binom{k}{2} = \binom{m}{2}\binom{m+n-2}{r-2}$ leads to Eq. (F.7) given in Box I. Combining Eqs. (F.2) and (F.7) with the identity $\sum_{j=0}^{k} \binom{k}{j}\binom{j}{2} = 2^{k-2}\binom{k}{2}$, we obtain

$$\begin{aligned}
&\operatorname{tr}\left(\left(\lim_{N\to\infty} P^{(k,k)}\right)\left(\lim_{N\to\infty} N^2 \frac{d}{dN} P^{(k,k)}\right)\right) \\
&= \frac{1}{4^k}\left(\sum_{j=0}^{k}\binom{k}{j}\left(r^{-1}\binom{j}{2} + (1-r)^{-1}\binom{k-j}{2}\right)\right) \\
&\quad \times \sum_{i=0}^{k}\left(\binom{k-2}{i-2} + \binom{k-2}{i}\right) \\
&= \frac{\binom{k}{2}}{8r(1-r)}. \quad (\text{F.8})
\end{aligned}$$

Now substituting Eqs. (F.2) and (F.8) into Eq. (F.3) yields

$$F_{k^1,\ell^1}^{(k)} = \frac{8r(1-r)}{\binom{k}{2}} \frac{1}{2^k}\binom{k}{\ell^1}, \quad (\text{F.9})$$

recovering Eq. (6.14b).

Finally, we compute $\Psi^{(k-1)} = F^{(k)}\left(\lim_{N\to\infty} N P^{(k,k-1)}\right)$. We evaluate $\lim_{N\to\infty} N P^{(k,k-1)}$ in similar fashion to $\lim_{N\to\infty} N^2 \frac{d}{dN} P^{(k,k)}$, leading to

$$\lim_{N\to\infty} N P_{k^1,\ell^1}^{(k,k-1)} = \frac{1}{2^k}\left(r^{-1}\binom{k^1}{2} + (1-r)^{-1}\binom{k - k^1}{2}\right)\binom{k-1}{\ell^1}. \quad (\text{F.10})$$

Combining with Eq. (F.9) and $\sum_{j=0}^{k} \binom{k}{j}\binom{j}{2} = 2^{k-2}\binom{k}{2}$, we obtain

$$\Psi_{k^1,\ell^1}^{(k-1)} = \frac{1}{2^{k-1}}\binom{k-1}{\ell^1}, \quad (\text{F.11})$$

recovering Eq. (6.14c). We observe that each row of $\Psi^{(k-1)}$ contains the probabilities in the binomial distribution $\mathrm{Binom}(k - 1, 1/2)$. The (informal) intuition for this finding is that, for finite $N$,

$$\left(N^{-1}\left(I - P^{(k,k)}\right)^{-1}\right)\left(N P^{(k,k-1)}\right) = \sum_{t=0}^{\infty}\left(P^{(k,k)}\right)^t P^{(k,k-1)}. \quad (\text{F.12})$$

The $(k^1, \ell^1)$ entry, on each side of Eq. (F.12), is the probability that, from an initial state with $k^1$ out of $k$ ancestral lineages in female-inherited sites, there are $\ell^1$ lineages in female-inherited sites at the moment when two of the $k$ lineages meet. As $N \to \infty$, the expected time to a meeting of lineages diverges, giving $\ell^1$ time to forget the initial condition, $k^1$, and relax to the stationary distribution, which is $\mathrm{Binom}(k, 1/2)$ by Eq. (F.2). Then, after a pair of lineages meet and $k - 1$ distinct lineages remain, the number of these lineages in female-inherited sites is distributed as $\mathrm{Binom}(k - 1, 1/2)$.

Box I.

$$\lim_{N \to \infty} N^2 \frac{d}{dN} P^{(k,k)}_{k^1,\ell^1}$$
$$= \sum_{\ell^1_F + \ell^1_M = \ell^1} \begin{pmatrix} \lim_{N \to \infty} N^2 \frac{d}{dN} \mathbb{P}_F\left[(\ell^1_F, k^1 - \ell^1_F) \mid k^1\right] \lim_{N \to \infty} \mathbb{P}_M\left[(\ell^1_M, k^2 - \ell^1_M) \mid k^2\right] \\ + \lim_{N \to \infty} \mathbb{P}_F\left[(\ell^1_F, k^1 - \ell^1_F) \mid k^1\right] \lim_{N \to \infty} N^2 \frac{d}{dN} \mathbb{P}_M\left[(\ell^1_M, k^2 - \ell^1_M) \mid k^2\right] \end{pmatrix}$$
$$= \frac{1}{2^k} \sum_{\ell^1_F + \ell^1_M = \ell^1} \begin{pmatrix} r^{-1} \binom{k^1}{\ell^1_F}\binom{k^2}{\ell^1_M}\left(\binom{\ell^1_F}{2} + \binom{k^1 - \ell^1_F}{2}\right) \\ + (1-r)^{-1} \binom{k^1}{\ell^1_F}\binom{k^2}{\ell^1_M}\left(\binom{\ell^1_M}{2} + \binom{k^2 - \ell^1_M}{2}\right) \end{pmatrix}$$
$$= \frac{1}{2^k}\left(r^{-1}\binom{k^1}{2} + (1-r)^{-1}\binom{k - k^1}{2}\right)\left(\binom{k-2}{\ell^1 - 2} + \binom{k-2}{\ell^1}\right). \tag{F.7}$$